\documentclass[times,twocolumn,final]{elsarticle}

\usepackage{medima}

\usepackage{amsmath,amssymb,amsfonts}
\usepackage{algorithmic}
\usepackage{graphicx}
\usepackage{float}
\usepackage{adjustbox}
\usepackage{stfloats}
\usepackage[justification=justified]{caption}
\usepackage{textcomp}
\usepackage[utf8]{inputenc}
\usepackage{longtable}
\usepackage{url}
\usepackage{xr}
\usepackage{multirow}
\usepackage{xcolor}
\usepackage{comment}
\usepackage[colorlinks]{hyperref}
\hypersetup{
  colorlinks,
  citecolor=teal,
  linkcolor=Blue,
  urlcolor=Blue}

\makeatletter
\def\ps@pprintTitle{%
  \let\@oddhead\@empty
  \let\@evenhead\@empty
  \let\@oddfoot\@empty
  \let\@evenfoot\@oddfoot
}

\def\stmtitleheader{}

\def\ps@headings{}

\def\ps@MEDIMATitle{}

\def\@oddhead{}%
  \let\@evenhead\@empty
  \def\@oddfoot{%
      \ifpreprint%
        \ifnopreprintline\relax\else%
             Preprint submitted to \ifx\@journal\@empty%
               Elsevier%
           \else\@journal\fi
        \fi%
      \fi%
  }
  \let\@evenfoot\@oddfoot

\renewenvironment{abstract}{%
  \global\setbox\absbox=\vtop\bgroup%
  \hsize=\textwidth%
  \leftskip=.335\textwidth
  \noindent\unskip\ignorespaces%
  }
 {\mbox{}\newline%
  \mbox{~}\hfill \ %
  \egroup%
 }%
 
 \def\articleinfobox{%
    \parbox[t]{.25\textwidth}{%
    \vspace*{0pt}%
    \fontsize{8pt}{10pt}\selectfont%
    \textit{}\\
    \ifx\@received\@empty\relax
    \else
    \receivedhead~\@received\\
    \fi
    \ifx\@finalform\@empty\relax
    \else
    \finalformhead~\@finalform\\
    \fi
    \ifx\@accepted\@empty\relax
    \else
    \acceptedhead~\@accepted\\
    \fi
    \ifx\@availableonline\@empty\relax
    \else
    \availableonlinehead~\@availableonline
    \fi
    \ifx\@communicated\@empty\relax
    \\[-2pt]
    \else
    \par
    \vspace*{10pt}
    \communicatedhead~\@communicated
    \vspace*{12pt}
    \fi
    \hrule width 0.285\textwidth
    \vspace*{1pc}
    \unhbox\keybox}
}

\long\def\pprintMaketitle{\MaketitleBox}%

\makeatletter
\def\widebreve{\mathpalette\wide@breve}
\def\wide@breve#1#2{\sbox\z@{$#1#2$}%
     \mathop{\vbox{\m@th\ialign{##\crcr
\kern0.08em\brevefill#1{0.8\wd\z@}\crcr\noalign{\nointerlineskip}%
                    $\hss#1#2\hss$\crcr}}}\limits}
\def\brevefill#1#2{$\m@th\sbox\tw@{$#1($}%
  \hss\resizebox{#2}{\wd\tw@}{\rotatebox[origin=c]{90}{\upshape(}}\hss$}
\makeatletter

\makeatletter
\def\widecaptionbreve#1{\breve{#1}}
\makeatletter

\def\SB#1{\textsubscript{#1}}
\def\BibTeX{{\rm B\kern-.05em{\sc i\kern-.025em b}\kern-.08em
    T\kern-.1667em\lower.7ex\hbox{E}\kern-.125emX}}

\begin{document}

\begin{frontmatter}
\title{Adaptive Diffusion Priors for Accelerated MRI Reconstruction}
\author[1,2,3,o]{Alper G\"{u}ng\"{o}r}
\author[1,2,o]{Salman UH Dar}
\author[1,2,4,o]{\c{S}aban \"{O}zt\"{u}rk}
\author[1,2]{Yilmaz Korkmaz}
\author[1,2]{Hasan A Bedel}
\author[1,2]{Gokberk Elmas}
\author[1,2]{\\Muzaffer Ozbey}
\author[1,2,5]{Tolga \c{C}ukur\corref{cor1}}
\cortext[cor1]{Corresponding author, 
  e-mail: cukur@ee.bilkent.edu.tr}
\address[1]{Department of Electrical and Electronics Engineering, Bilkent University, Ankara 06800, Turkey}
\address[2]{National Magnetic Resonance Research Center (UMRAM), Bilkent University, Ankara 06800, Turkey}
\address[3]{ASELSAN Research Center, Ankara 06200, Turkey}
\address[4]{Department of Electrical and Electronics Engineering, Amasya University, Amasya 05100, Turkey}
\address[5]{Neuroscience Program, Bilkent University, Ankara 06800, Turkey}
\address[o]{denotes equal contribution}


\begin{abstract}
Deep MRI reconstruction is commonly performed with conditional models that de-alias undersampled acquisitions to recover images consistent with fully-sampled data. Since conditional models are trained with knowledge of the imaging operator, they can show poor generalization across variable operators. Unconditional models instead learn generative image priors decoupled from the operator to improve reliability against domain shifts related to the imaging operator. Recent diffusion models are particularly promising given their high sample fidelity. Nevertheless, inference with a static image prior can perform suboptimally. Here we propose the first adaptive diffusion prior for MRI reconstruction, AdaDiff, to improve performance and reliability against domain shifts. AdaDiff leverages an efficient diffusion prior trained via adversarial mapping over large reverse diffusion steps. A two-phase reconstruction is executed following training: a rapid-diffusion phase that produces an initial reconstruction with the trained prior, and an adaptation phase that further refines the result by updating the prior to minimize data-consistency loss. Demonstrations on multi-contrast brain MRI clearly indicate that AdaDiff outperforms competing conditional and unconditional methods under domain shifts, and achieves superior or on par within-domain performance. 
\end{abstract}


\end{frontmatter}

\section{Introduction}
Magnetic resonance imaging (MRI) is a preferred modality in diagnostic applications due to its exceptional soft-tissue contrast, yet canonically long exams hinder its clinical use. A fundamental solution is to shorten scan times by undersampling k-space acquisitions and solve an ill-posed inverse problem to reconstruct images \citep{Lustig2007,gu2021compressed}. In recent years, deep learning methods have become a gold standard in MRI reconstruction, given their ability to solve complex inverse problems based on data-driven priors \citep{Dong2020spm,Yang2016,Wang2016,Hammernik2017,Schlemper2017,Dar2017,Kwon2017,yaman2021zero}. Many proposed methods are based on conditional models that process undersampled acquisitions provided as input to recover output images that are consistent with fully-sampled acquisitions \citep{Zhu2018,MoDl,Hyun2018,lee2018deep,KnollGeneralization,Yoon2018,ChulYe2018,Quan2018c,Yu2018c,Adler2018,guo2021over}. This conditional mapping can be learned explicitly from a training dataset of paired undersampled and fully-sampled acquisitions \citep{rgan,Mardani2019b,Biswas2019,Wang2019,ADMM-CSNET,KikiNet,Primal_dual,Conv_recur,Hosseini2020b}. To alleviate requirements on training data, the mapping can also be learned implicitly on undersampled acquisitions via self-supervision \citep{Tamir2019,Wang2020self,Cole2020,yaman2020,Huang2019self,Liu2020,aggarwal2020} or cycle-consistency approaches \citep{Quan2018c,oh2020,lei2020,chung2020progressive,lu2021two}. Regardless of the learning strategy, conditional models capture a de-aliasing prior to suppress undersampling artifacts, so they have explicit knowledge of the imaging operator that reflects the choice of sampling patterns and coil sensitivities for acceleration \citep{Polakjointvvn2020,feng2021donet,kustner2020cinenet,Variatonal_end2end}. Deep reconstruction models are commonly trained based on a relatively standardized imaging operator to help maximize performance \citep{fastmri}. However, in the testing stage, an end user might have to prescribe spontaneous changes to the imaging operator (e.g. changes in acceleration rate, sampling density or number of coils) in order to meet practical considerations on image quality or scan time. Since the imaging operator inherently determines the characteristics of aliasing artifacts in undersampled acquisitions, such domain shifts in the operator can compromise reconstruction performance and necessitate re-training of conditional models \citep{Knoll2019inverseGANs,liu2021universal}. To avoid potential losses in generalization performance, deep reconstruction models that are resilient against variations in the imaging operator between the training and test sets are direly needed.

\par
An alternative framework employs unconditional models that are not trained to perform the reconstruction task (i.e., mapping undersampled to fully-sampled data), but instead to capture generative image priors through auxiliary tasks such as additive noise removal \citep{PNPrizwan}, image autoencoding \citep{Konukoglu2019,Liu2020mrm,tezcan2022sampling} or image generation \citep{Knoll2019inverseGANs,Darestani2021,pixelcnnrecon,korkmaz2022unsupervised,elmas2022federated}. Image priors are only combined with the imaging operator during inference, so they improve generalization against variable operators as they are agnostic to undersampling \citep{Konukoglu2019,Knoll2019inverseGANs}. Adversarial priors are particularly prominent as they offer elevated sensitivity to detailed tissue structure \citep{Knoll2019inverseGANs,korkmaz2022unsupervised}, but they might manifest poor diversity in generated image samples \citep{DiffBeatsGAN}. As a promising surrogate, diffusion models enhance sample diversity while maintaining comparable sample quality \citep{DDPM}. Recent studies have reported remarkable reconstructions with alternated projections through diffusion priors and through the imaging operator to enforce consistency to acquired data \citep{jalaln2021nips, chung2022media,song2022,chung2022cvpr,luo2022uncertainty,xie2022kspace,peng2022}. Still, static diffusion priors can limit model performance under domain shifts in the MR image distribution, which can result from changes in the pulse sequence or scanner settings. 

Here we introduce a novel diffusion-based method, AdaDiff, to improve performance and reliability against domain shifts in accelerated MRI reconstruction. AdaDiff learns an unconditional diffusion prior for high-fidelity image generation (Fig. \ref{fig:training}), and adapts the diffusion prior during inference for enhanced performance (Fig. \ref{fig:inference}). Vanilla diffusion models generate images through a long sequence of inference steps, resulting in prolonged image sampling \citep{DDPM}. We instead propose a diffusion model based on an adversarial mapper to generate images in few, large reverse diffusion steps for a notable speed up in image sampling. A two-phase reconstruction is then employed during inference: a rapid-diffusion phase that produces an initial reconstruction by fast image sampling based on the trained prior, and an adaptation phase that produces a refined reconstruction by updating the prior to minimize data-consistency loss. The adversarial diffusion prior enables AdaDiff to reconstruct high-quality images in fewer inference iterations compared to static, untrained or non-adversarial diffusion priors.

The proposed method is demonstrated for reconstruction of multiple contrasts in brain MRI based on a unified model trained on mixed contrasts. Experiments are reported for within-domain cases with matched imaging operator and image distribution between training-test sets, and cross-domain cases with a domain shift in the operator or the MR image distribution. Comparisons are provided against state-of-the-art traditional, conditional and unconditional deep models. In general, AdaDiff achieves superior or on par performance in within-domain cases, and outperforms competing methods in cross-domain cases. Code to implement AdaDiff is available at {\small \url{https://github.com/icon-lab/AdaDiff}}.

\subsubsection*{\textbf{Contributions}}
\begin{itemize}
    \item To our knowledge, AdaDiff is the first prior adaptation method based on diffusion models in literature for accelerated MRI reconstruction.
    \item The proposed method leverages a rapid diffusion process with an adversarial mapper for efficient sampling from the diffusion prior.
    \item Inference adaptation is performed on the trained diffusion prior to improve performance and reliability against domain shifts. 
\end{itemize}

\section{Related Work}
Unconditional models that decouple the image prior from the imaging operator promise enhanced generalization in MRI reconstruction \citep{Knoll2019inverseGANs,Konukoglu2019,Liu2020mrm,korkmaz2022unsupervised,PNPrizwan,Darestani2021,pixelcnnrecon}. In this generative modeling (GM) framework, image priors are typically learned to capture information regarding the distribution of high-quality MRI data \citep{Knoll2019inverseGANs,Konukoglu2019,Liu2020mrm,korkmaz2022unsupervised,elmas2022federated,pixelcnnrecon}. A common approach rests on generative adversarial networks (GANs) that indirectly characterize the data distribution \citep{Knoll2019inverseGANs,Liu2020mrm,korkmaz2022unsupervised,elmas2022federated}. Despite the realism of generated image samples, GAN-based methods can be susceptible to low representational diversity that can hamper reconstruction performance. 

A recent class of GMs based on diffusion promise enhanced representational diversity over GANs. Diffusion models use a multi-step process to gradually transform Gaussian noise into image samples. Unlike GANs, they directly characterize correlates of the data distribution (e.g., derivative or lower bound of log-likelihood). The learned priors are coupled with the imaging operator at time of inference \citep{jalaln2021nips, chung2022media,song2022,chung2022cvpr,luo2022uncertainty,peng2022}. Reconstruction can then be performed via repeated projections through the diffusion prior and the operator. Projections through the diffusion prior involve generation of image samples. For instance, \cite{jalaln2021nips,luo2022uncertainty} proposed sampling with score-based functions and Langevin dynamics; \cite{chung2022media,song2022} used a predictor to solve a stochastic differential equation followed by Langevin sampling. High image quality has typically been reported with diffusion-based MRI reconstruction \citep{peng2022}.

Despite their prowess, diffusion-based methods are not without limitation. Vanilla diffusion models use hundreds of reverse steps for image generation \citep{jalaln2021nips}, elevating computational burden. \cite{peng2022} considered rescaling the diffusion step size during inference to accelerate image sampling, but this can potentially reduce the accuracy of reverse diffusion steps \citep{DDPM}. \cite{chung2022cvpr} proposed to obtain an initial reconstruction via a separate method, and then initiate the reverse diffusion process with this initial reconstruction for faster inference. While promising, this approach involves implementation of a second reconstruction method. Furthermore, existing diffusion methods learn a static prior that is kept fixed during inference. In turn, a trained prior might be rendered suboptimal by domain shifts in the image distribution between the training-test sets \citep{Knoll2019inverseGANs}.  

Here we propose an adaptive diffusion prior, AdaDiff, for MRI reconstruction. AdaDiff differs from recent GM-based reconstruction methods in several key aspects. Unlike GAN-based methods that use adversarial learning for single-shot mapping from noise variables onto images, AdaDiff is based on a multi-step diffusion process to improve fidelity in generated image samples. Unlike methods based on static diffusion priors, AdaDiff performs subject-specific adaptation of its prior during inference to increase conformity of its prior to the distribution of the test data. Finally, unlike diffusion methods based on a long-chain of sampling steps, AdaDiff performs diffusion modeling in few large steps implemented via adversarial mapping for enhanced reliability.

\begin{figure*}[!h]
\centering
\includegraphics[width=0.7\textwidth]{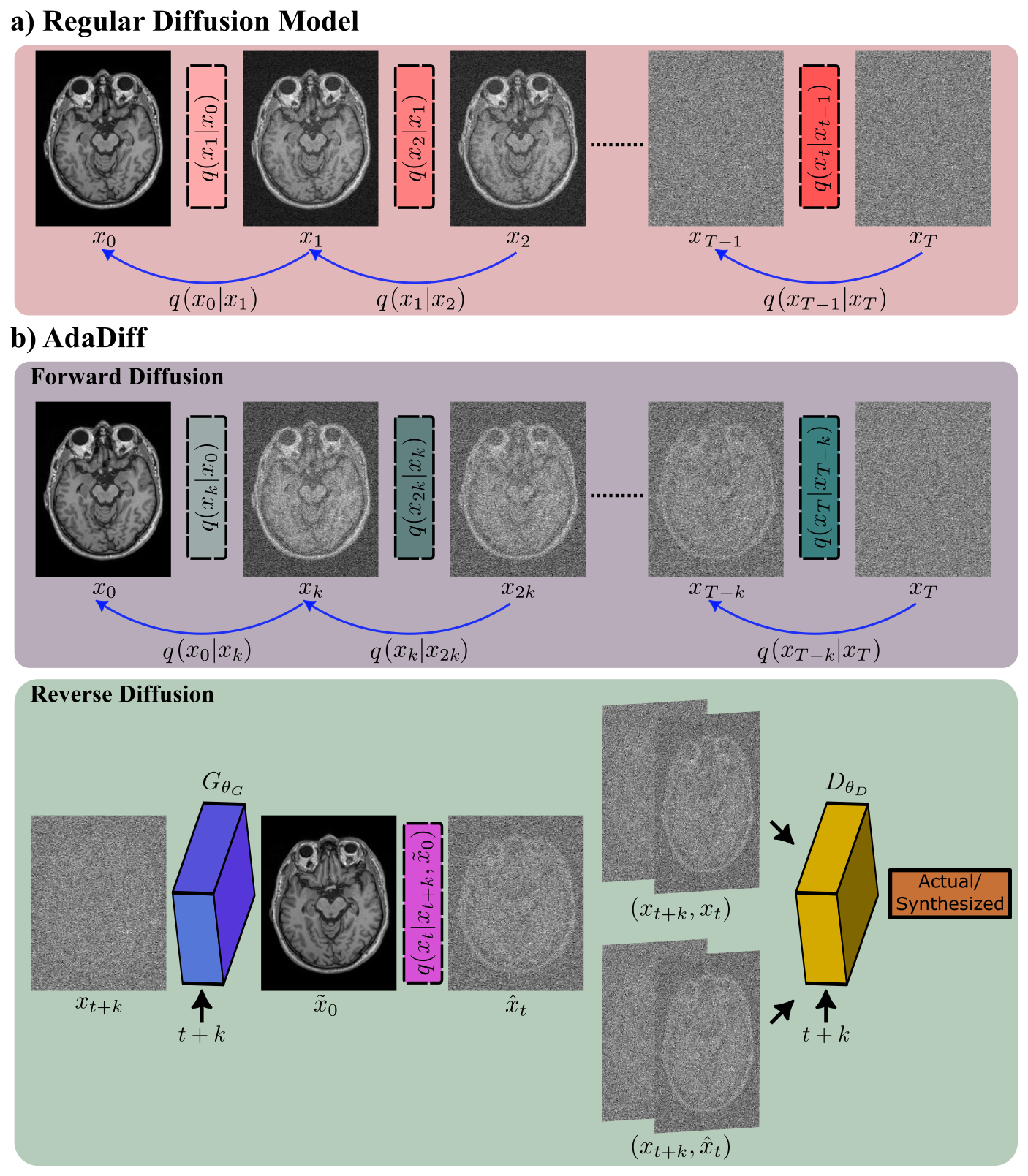}
\captionsetup{justification   = justified,singlelinecheck = false}
\caption{\textbf{a)} Diffusion models generate an actual image ($x_0$) starting from isotropic Gaussian noise ($x_T$) through a gradual process with forward and reverse steps. In a forward step, scaled Gaussian noise is added onto the previous sample $x_{t-1}$ to obtain a noisier sample $x_t$ (Eq. \ref{eq:diff_forward}). In a reverse step, additive noise on $x_{t+1}$ is suppressed to obtain $x_t$. This reverse mapping is parameterized as projection through a neural network, $p_{\theta}\left( x_{t}|x_{t+1} \right)$. Vanilla diffusion models use small step sizes to ensure approximate normality of the reverse transition probability $q\left( x_{t}|x_{t+1} \right)$, resulting in prolonged sampling. \textbf{b)} AdaDiff leverages rapid diffusion with a large step size $k$ to transform between $x_0$ and $x_T$ in few steps (Eq. \ref{eq:our_forward}). Because the added noise in each step is scaled up to account for large step size, the normality assumption for the reverse transition probability $q\left( x_{t}|x_{t+k} \right)$ breaks down. To address this issue, AdaDiff employs an adversarial mapper that implicitly characterizes the distribution of the reverse diffusion steps. The generator estimates denoised image samples (Eq. \ref{eq:G-loss1}), whereas the discriminator distinguishes actual samples based on the forward diffusion process from synthesized samples produced by the generator (Eq. \ref{eq:D-loss1}). \vspace{-0.4cm}}
\label{fig:training}
\end{figure*}

\section{Theory}

\subsection{MRI Reconstruction}
Accelerated MRI entails recovery of a subject's MR image $x$ from undersampled k-space acquisitions $y$:
\begin{equation}
\label{eq:acceleratedmri}
Ax = y
\end{equation}
where $A=\Omega \mathcal{F} B$ is the imaging operator that captures the influence of the k-space undersampling pattern ($\Omega$) and coil sensitivities ($B$), and $\mathcal{F}$ denotes Fourier transform. Since the inverse problem in Eq. \ref{eq:acceleratedmri} is ill-posed, prior information is typically incorporated to obtain a reconstruction $\widebreve{x}$:
\begin{equation}
\label{eq:regacceleratedmri}
 \widebreve{x} = \underset{x}{\operatorname{min}} \left\|Ax-y\right\|_{2} + R(x,y)
\end{equation} 
where $R(x,y)$ denotes the regularization term that enforces the prior. Given a training set of MRI data, conditional models capture a de-aliasing prior often conditioned on the inverse Fourier transform of undersampled data as input $R(x|\mathcal{F}^{-1}(y))$. Instead, unconditional models capture a generative image prior agnostic to undersampling, $R(x)$. Since image priors are not tied to specific imaging operators, they promise improved reliability against domain shifts in the operator.

\subsection{Diffusion Models}
Diffusion models are likelihood-based GMs that express image generation as a temporal Markov process (Fig. \ref{fig:training}a). Modeling involves forward and reverse processes that conventionally comprise hundreds of steps \citep{DDPM}. The forward process adds a small amount of isotropic Gaussian noise $z\sim \mathcal{N}\left( 0,\mathrm{I} \right)$ in each step to modify an actual image \(x_0\sim q\left( x_{0} \right)\) at time step $0$, and produce a sequence of noisy samples \(x_{1:T}\) where $T$ is the final time step. At step $t$, the relationship between $x_t$ and $x_{t-1}$, and the corresponding conditional distribution $q\left( x_{t}|x_{t-1} \right)$ can be described as follows:
\begin{eqnarray}
\label{eq:diff_forward}
\displaystyle x_{t}&=&\sqrt{1-\beta_{t}} x_{t-1}+\sqrt{\beta_{t}} z, \\
q\left( x_{t}|x_{t-1} \right) &=&\mathcal{N}\left( x_{t}; \sqrt{1-\beta_{t}}x_{t-1},\beta_{t}\mathrm{I} \right)
\end{eqnarray}
where $\beta_{t}$ is the noise scaling. After a large number of forward steps, $x_{t}$ approaches an isotropic Gaussian sample. 

The reverse process gradually removes the added noise in $x_T$ to recollect $x_0$. Diffusion models operationalize each reverse step as $x_{t}=f_{\theta}(x_{t+1})$, where $f_{\theta}$ denotes projection through a network with parameters $\theta$. The network can be trained to minimize a lower bound on the negative log-likelihood:
\begin{eqnarray}
\label{eq:likelihood}
L_{\text{lb}} &=& \sum_{t=0} ^{T-1} D_{KL}\left( q\left( x_{t}|x_{t+1} \right)\parallel p_{\theta}\left( x_{t}|x_{t+1} \right) \right)
\end{eqnarray}
where $p_{\theta}\left( x_{t}|x_{t+1} \right)$ denotes the parametrization of $q\left( x_{t}|x_{t+1} \right)$, and $D_{KL}$ is the Kullback-Leibler (KL) divergence. Note that $q\left( x_{t}|x_{t+1}\right)$ is generally unknown, so Eq. \ref{eq:likelihood} cannot be evaluated and an alternative formulation is adopted: 
\begin{eqnarray}
\label{eq:likelihood2}
L_{\text{lb}} &=& -log p_{\theta}\left( x_{0}|x_{1} \right) + \nonumber \\ && \sum_{t=1} ^{T-1}  D_{KL}\left( q\left( x_{t}|x_{t+1},x_{0} \right)\parallel p_{\theta}\left( x_{t}|x_{t+1} \right) \right) 
\end{eqnarray}
where the auxiliary distribution $q\left( x_{t}|x_{t+1},x_{0} \right)$ has a closed-form expression \citep{DDPM}. For small step sizes, $q\left( x_{t}|x_{t+1},x_{0}\right) \approx q\left(x_{t}|x_{t+1} \right)$, so a generator network can learn the desired mapping by minimizing Eq. \ref{eq:likelihood2}. A common approach to minimize $L_{\text{lb}}$ is to predict the additive noise $z$ given $x_{t+1}$ and $x_0$ as input \citep{DDPM}:
\begin{equation}
\min_{\theta} \mathbb{E} \left[ \left\|\sqrt{\beta_{t+1}}z - (x_{t+1}-\sqrt{1-\beta_{t+1}} f_{\theta}(x_{t+1})\right\|_{2} \right]
\end{equation}
where $\mathbb{E}$ is expectation of the difference norm between the scaled noise added onto $x_{t}$ in the forward step and the noise predicted by the network in the reverse step. 

A trained diffusion model can be used to generate random image samples during inference. Starting from a noise image $x_{T} \sim \mathcal{N}\left( 0,\mathrm{I} \right)$, hundreds of reverse steps are performed using $p_{\theta}\left( x_{t}|x_{t+1} \right)$ to obtain an image sample $x_0$. For MRI reconstruction, reverse diffusion projections are interleaved with data-consistency projections to align the generated image sample with the acquired k-space data for each subject \citep{luo2022uncertainty,peng2022}. However, recent reconstruction methods use static diffusion priors that can potentially elicit suboptimal performance.

\subsection{AdaDiff}
\begin{figure*}[!h]
\centering
\includegraphics[width=0.7\textwidth]{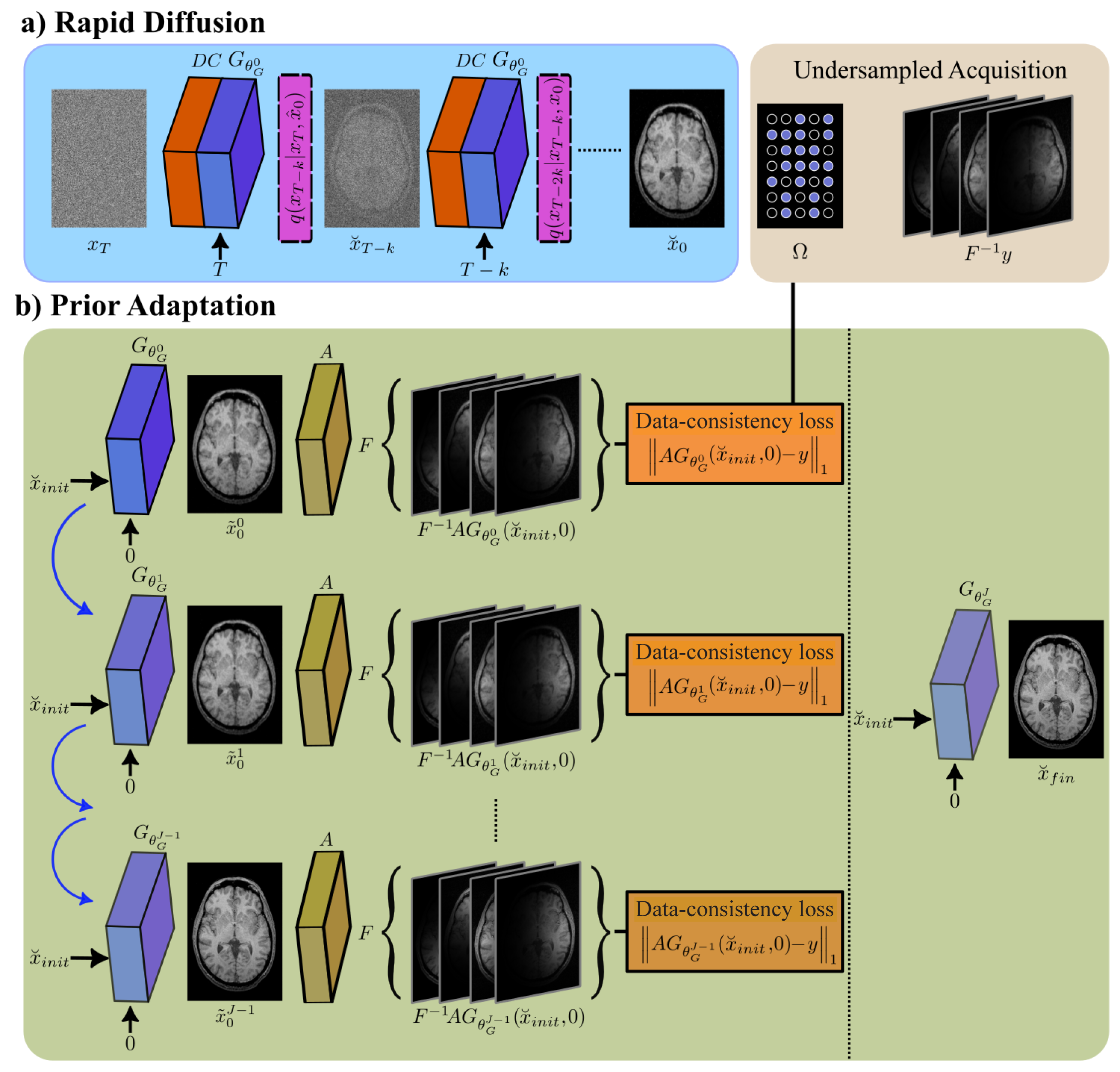}
\captionsetup{justification   = justified,singlelinecheck = false}
\caption{AdaDiff employs a two-phase reconstruction given a learned diffusion prior with a trained generator $G_{\theta^0_{G}}$. \textbf{a)} The rapid diffusion phase calculates a fast, initial solution as a compromise between consistency with the learned prior and consistency with the imaging operator. Starting with a Gaussian noise sample $\widecaptionbreve{x}_T$, interleaved projections are performed through data-consistency (DC) blocks (Eq. \ref{eq:datacons}) and reverse diffusion steps (Eq. \ref{eq:optimizationx0pred}). The sample at time step 0 is taken as the initial reconstruction, $\widecaptionbreve{x}_0 = \widecaptionbreve{x}_{init}$. \textbf{b)} The adaptation phase refines the diffusion prior per test subject to further improve the initial reconstruction. To do this, the generator parameters ($\theta_{G}$) are iteratively optimized to minimize a data-consistency loss (Eq. \ref{optimizationequationadaptation}). At the $j^{th}$ iteration, the generator receives the initial reconstruction $\widecaptionbreve{x}_{init}$ to synthesize a coil-combined image $\tilde{x}_0^j$. Synthetic multi-coil images are obtained by projecting $\tilde{x}_0^j$ through the imaging operator $A$ that encapsulates estimated coil sensitivities and undersampling in k-space with the subject's prescribed sampling mask $\Omega$. The data-consistency loss is taken as the difference between synthesized and acquired data in k-space. The generator output at the end of $J$ iterations is taken as the final reconstruction $\widecaptionbreve{x}_{fin}$.}
\label{fig:inference}
\end{figure*}

Here we propose to perform prior adaptation on a diffusion model for enhanced performance in MRI reconstruction. AdaDiff is trained to efficiently generate high-quality image samples via a rapid diffusion process with substantially fewer steps than typically used (Fig. \ref{fig:training}b). Vanilla diffusion models assume approximate normality for reverse transition probabilities $q\left( x_{t}|x_{t+1}\right)$, but this assumption breaks down with increasing step size \citep{DDPM,DiffNvidia}. To improve accuracy, we propose an adversarial mapper to parametrize the reverse diffusion steps as inspired by a recent study on natural image synthesis \citep{DiffNvidia}. Given a trained diffusion prior, a two-phase reconstruction is employed to recover a subject's images during inference (Fig. \ref{fig:inference}). In the rapid diffusion phase, an initial reconstruction is obtained via interleaved reverse diffusion and data-consistency projections. In the prior adaptation phase, the prior is combined with the imaging operator to evaluate a data-consistency loss based on the difference between generated and measured k-space data. The parameters of the prior are iteratively updated to minimize the data-consistency loss. The image generated by the adapted prior is taken as the final reconstruction.

\subsubsection{Training of the Prior}
Vanilla diffusion models prescribe small step sizes to approximate $q\left( x_{t}|x_{t+1} \right)$ with an auxiliary Gaussian distribution, necessitating computationally-intensive inference with a large number of diffusion steps. We instead adopt a rapid adversarial diffusion model with a large step size of $k$ as described in \citep{DiffNvidia}:
\begin{eqnarray}
\label{eq:our_forward}
\displaystyle x_{t}&=&\sqrt{1-\gamma_{t}}x_{t-k}+\sqrt{\gamma_{t}} z, \\
q\left( x_{t}|x_{t-k} \right) &=&\mathcal{N}\left( x_{t}; \sqrt{1-\gamma_{t}}x_{t-k},\gamma_{t}\mathrm{I} \right)
\end{eqnarray}
where the noise variance $\gamma_t$ has to be greater than $\beta_t$ to compensate for large $k$. Because the normality assumption does not hold for $ q\left(x_{t}|x_{t+k}\right)$, the original likelihood formulation must be considered for the lowerbound:
\begin{align}
\label{eq:VLB}
L_{\text{lb}} = \sum_{\substack{t=rk \\ r=[0,..,T/k-1]}}   D_{KL}\left( q\left( x_{t}|x_{t+k} \right)\parallel p_{\theta}\left( x_{t}|x_{t+k} \right) \right)
\end{align}
There is no closed-form expression for $q\left(x_{t}|x_{t+k}\right)$, so we introduce an adversarial mapper to implicitly capture the conditional distribution for the reverse diffusion steps. A generator $G_{\theta_{G}}$ is used to parametrize a sampling distribution $p_{\theta_G}\left( x_{t}|x_{t+k} \right)$ that synthesizes $\hat{x}_t$. Meanwhile, a discriminator $D_{\theta_{D}}$ differentiates between synthetic samples ($\hat{x}_t$) drawn from $p_{\theta_G}\left( x_{t}|x_{t+k} \right)$ and actual samples ($x_t$) drawn from the true denoising distribution $q\left( x_{t}|x_{t+k} \right)$. Both $G_{\theta_{G}}$ and $D_{\theta_{D}}$ receive time index $t+k$ as input. In this framework, the discriminator is trained to minimize an adversarial loss \citep{Dar2019} coupled with a gradient penalty to improve learning \citep{pmlr-v80-mescheder18a}:
\begin{align}
\label{eq:D-loss1}
L_{D} = \sum_{t\ge 0}^{}\mathbb{E}_{q\left( x_{t+k} \right)} [ \mathbb{E}_{q\left( x_{t}|x_{t+k}\right)}\left[ -\log\left( D_{\theta_{D}} \left( x_{t},x_{t+k},t+k \right)\right) \right] \notag \\
+\mathbb{E}_{p_{\theta_{G}}\left( x_{t}|x_{t+k} \right)}\left[ -\log\left( 1-D_{\theta_{D}}\left( \hat{x}_{t},x_{t+k},t+k \right) \right) \right]   \notag \\
+\mathbb{E}_{q\left( x_{t}|x_{t+k}\right)} [ \frac{1}{2} \left\| \nabla_{x_{t}} D_{\theta_D}( x_t,x_{t+k}, t)  \right\|_2 ] ]
\end{align}
To avoid saturation, the generator is trained accordingly to maximize the following loss function:
\begin{align}
\label{eq:G-loss1}
L_{G} = \sum_{t\ge 0}^{}\mathbb{E}_{q\left( x_{t+k}\right)p_{\theta_{G}}\left( x_{t}|x_{t+k} \right)}\left[ -\log\left( D_{\theta_{D}}\left( \hat{x}_{t},x_{t+k},t+k \right) \right) \right] 
\end{align}
While the first and third terms in Eq. \ref{eq:D-loss1} require sampling from the unknown $q(x_{t}|x_{t+k})$, an equivalent formulation can be derived in terms of the known forward distribution $q(x_{t+k}|x_{t})$:
\begin{align}
\label{eq:D-loss2}
\mathbb{E}_{q\left( x_{t+k} \right)q\left( x_{t}|x_{t+k} \right)} \approx
\mathbb{E}_{q\left( x_{0} \right)q\left( x_{t}|x_{0} \right)q\left( x_{t+k}|x_{t} \right)} =
\mathbb{E}_{q\left( x_{0},x_t \right)q\left( x_{t+k}|x_{t} \right)}
\end{align}
Meanwhile, Eq. \ref{eq:G-loss1} and the second term in Eq. \ref{eq:D-loss1} require sampling from the network parameterized distribution as $\hat{x}_t \sim p_{\theta_{G}}(x_{t}|x_{t+1})$. Although it is possible to use $G_{\theta_{G}}$ to predict $x_t$, estimates from an insufficiently trained generator at intermediate stages can yield suboptimal results. Thus, here we operationalize the sampling distribution based on the generator as:
\begin{align}
\label{eq:samplefromptheta}
p_{\theta_{G}}\left( x_{t}|x_{t+k} \right):= q\left( x_{t}|x_{t+k},\tilde{x}_{0} \right)
\end{align}
where the generator is used to estimate the denoised image sample at $t=0$ as $\tilde{x}_0 = G_{\theta_{G}} (x_{t+k},t+k)$. Assuming that $\tilde{x}_0$ is a reasonable estimate of $x_0$, $q\left( x_{t}|x_{t+k},\tilde{x}_{0} \right)$ can be shown to have a closed form expression $\mathcal{N}_{\theta_G}(\overline{\mu},\overline{\gamma})$ with:
\begin{eqnarray}
\label{eq:mu}
\overline{\mu}&=&\frac{\sqrt{\overline{\alpha}_{t}}\gamma_{t}}{1-\overline{\alpha}_{t+k}} \tilde{x}_{0}+\frac{\sqrt{\alpha_{t+k}}\left( 1-\overline{\alpha}_{t} \right)}{1-\overline{\alpha}_{t+k}}x_{t+k} \\
\label{eq:gamma}
\overline{\gamma}&=&\frac{1-\overline{\alpha}_{t}}{1-\overline{\alpha}_{t+k}}\gamma_{t+k}
\end{eqnarray}
where \( \alpha_{t}:=1-\gamma_{t}\) and \( \overline{\alpha}_{t}:=\prod_{\substack{\tau=rk \\r=[0,..,t/k]}}\alpha_{\tau}\) \citep{DDPM}. Since our proposed formulation does not involve a normality assumption on $q\left( x_{t}|x_{t+k} \right)$, it can improve accuracy of reverse diffusion mappings at large step sizes. 
Finally, the discriminator and generator losses can be expressed as:
\begin{align}
\label{eq:overall-loss}
L_{D} = \sum_{t\ge 0}^{} \Big( \mathbb{E}_{q\left( x_0,x_{t} \right)} \mathbb{E}_{q\left( x_{t+k}|x_{t}\right)}\left[ -\log\left( D_{\theta_{D}} \left( x_{t},x_{t+k},t+k \right)\right) \right]  \notag \\
+\mathbb{E}_{q\left( x_{t+k} \right)}\mathbb{E}_{\mathcal{N}_{\theta_{G}}\left( \overline{\mu},\overline{\gamma} \right)}\left[ -\log\left( 1-D_{\theta_{D}}\left( \hat{x}_{t},x_{t+k},t+k \right) \right) \right]   \notag \\
 +\mathbb{E}_{q\left( x_0,x_{t} \right)}\mathbb{E}_{q\left( x_{t+k}|x_{t}\right)} [ \frac{1}{2} \left\| \nabla_{x_{t}} D_{\theta_D}( x_t,x_{t+k}, t+k)  \right\|_2 ] \Big) \\
L_{G} = \sum_{t\ge 0}^{}\mathbb{E}_{q\left( x_{t+k} \right)}\mathbb{E}_{\mathcal{N}_{\theta_{G}}\left( \overline{\mu},\overline{\gamma} \right)}\left[ -\log\left( D_{\theta_{D}}\left( \hat{x}_{t},x_{t+k},t+k \right) \right) \right] 
 \end{align}

\subsubsection{Reconstruction with the Prior}
The diffusion prior is trained to capture the distribution of high-quality MR images so as to generate random image samples. However, these synthetic images do not correspond to a test subject as they are not informed about the imaging operator and the resultant acquired data \citep{Knoll2019inverseGANs}. Thus, reconstruction requires a solution at the intersection of the image sets spanned by the diffusion prior versus the operator. Here we propose a two-phase reconstruction with rapid diffusion and prior adaptation stages (Fig. \ref{fig:inference}). In rapid diffusion, an initial reconstruction is computed as a fast, compromise solution between the trained diffusion prior and the imaging operator. Note that the image sets for the diffusion prior and the operator can weakly intersect due to distributional shifts between training and test subjects. To improve performance, prior adaptation refines the reconstruction by updating the diffusion prior to better conform it to the distribution of individual test subjects. 

\textbf{Rapid diffusion}:
The rapid diffusion phase calculates an initial reconstruction ($\widebreve{x}_{init}$) that is a compromise solution between the image sets spanned by the imaging operator and the trained diffusion prior. This compromise solution can be obtained by alternating between data-consistency projections that sample images consistent with the operator, and reverse diffusion projections that sample images consistent with the trained diffusion prior \citep{jalaln2021nips}. Starting with $\widebreve{x}_T$ at time step $T$ randomly drawn from a Gaussian noise distribution, the two projections can be performed progressively across time steps. Given $\widebreve{x}_{t+k}$, the data-consistency projection at time step $t+k$ can be implemented as in \citep{peng2022}:
\begin{align}
\dot{x}_{t+k} = \widebreve{x}_{t+k} + A^H (y - A \widebreve{x}_{t+k})
\label{eq:datacons}
\end{align}
where $A^H$ denotes the Hermitian adjoint of $A$. The reverse diffusion projection can then be performed by sampling $\widebreve{x}_t$ from $q(x_{t}|x_{t+k},\tilde{x}_0)$ as described in Eq. \ref{eq:samplefromptheta}, where $x_{t+k}$ is taken as $\dot{x}_{t+k}$, and $\tilde{x}_0$ is computed via the generator as:
\begin{align}
 \tilde{x}_0= G_{\theta_{G}}(\dot{x}_{t+k},t+k)
\label{eq:optimizationx0pred}
\end{align}
Unlike conventional diffusion methods, AdaDiff leverages a rapid diffusion model with large step size. Thus, the initial reconstruction can be computed as $\widebreve{x}_{init} = \widebreve{x}_0$ in a few steps.

\begin{figure*}[!h]
\centering
\includegraphics[width=0.6\textwidth]{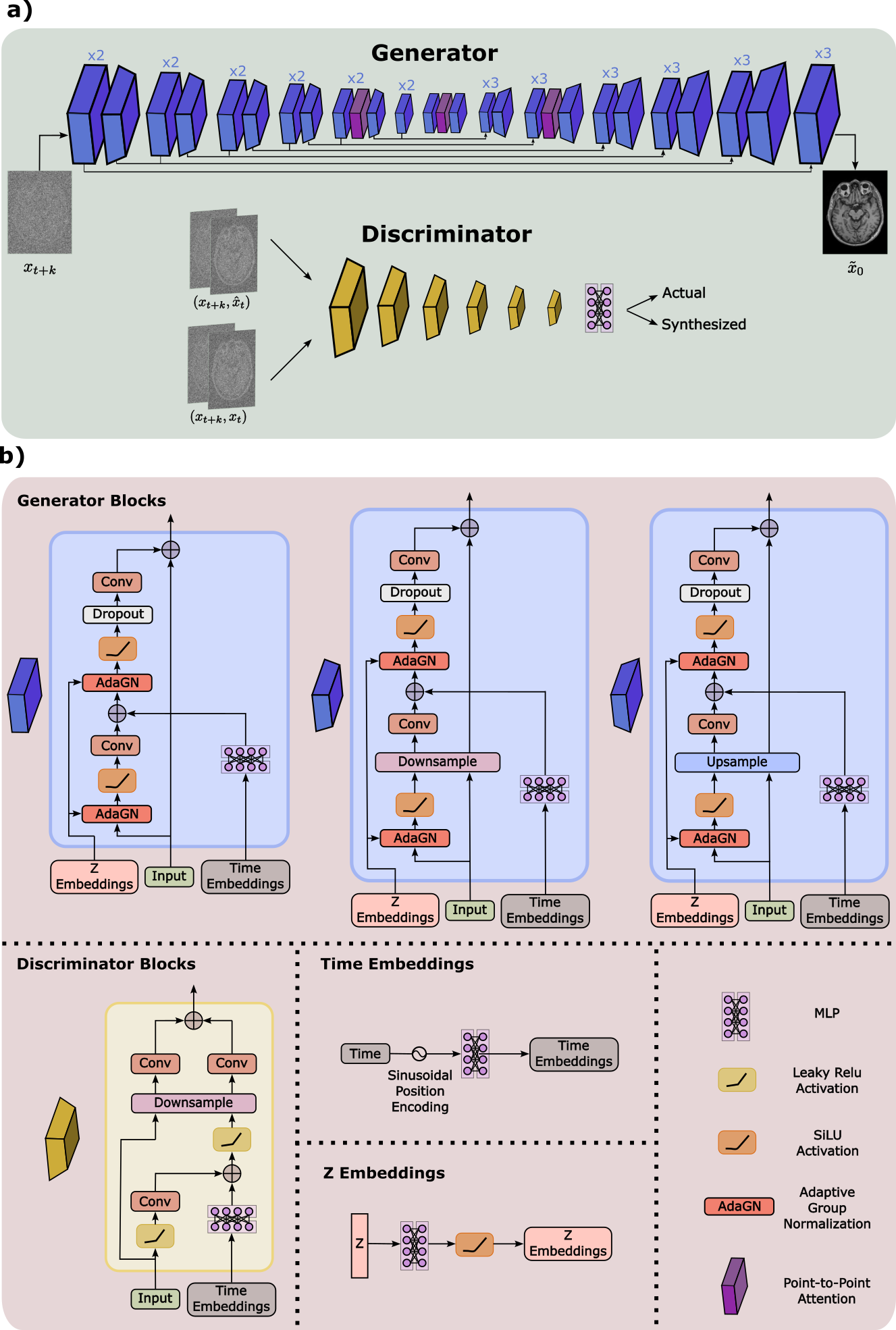}
\captionsetup{justification   = justified,singlelinecheck = false}
\caption{Architectural overview of AdaDiff. \textbf{a)} The generator processes the noisy image sample $x_{t+k}$ via an encoder-decoder architecture comprising residual (blue) and attentional (purple) blocks to predict the denoised image $\tilde{x}_0$. Long-range skip connections are used to enhance information flow across blocks. The discriminator processes a pair of noisy image samples, either ($x_{t+k},\hat{x}_{t}$) or ($x_{t+k},x_{t}$), with residual downsampling blocks to distinguish actual versus synthesized samples. An MLP block is used at the output layer. \textbf{b)} Detailed structure of the blocks used in adversarial network. Each discriminator block convolutionally encodes the input image samples and MLP encodes the time embeddings, followed by downsampling and convolutional filtering. Three types of residual blocks are used in the generator based on feature map resolution: flat blocks (rectangular), downsampling blocks (trapezoidal), upsampling blocks (inverted trapezoidal). Each block convolutionally encodes the input image and MLP encodes the time embeddings, followed by convolutional filtering and/or downsampling/upsampling. Feature maps are subjected to adaptive normalization based on Z (i.e., latent variable) embeddings. Time embeddings are obtained by processing sinusoidal encoding of time step with an MLP. Z embeddings are obtained by processing a random latent vector $z$ with an MLP. \vspace{-0.4cm}}
\label{fig:architecture}
\end{figure*}

\textbf{Prior adaptation}:
Taking the initial reconstruction $\widebreve{x}_{init}$ as input, the adaptation phase further refines the diffusion prior to improve the reconstruction. To do this, the generator parameters ($\theta_{G}$) are fine-tuned to minimize a data-consistency loss between synthesized and acquired k-space data:
\begin{align}
\theta_{G}^* = \min_{\theta_{G}}\left\|A G_{\theta_{G}}(\widebreve{x}_{init},0)-y\right\|_{1}
\label{optimizationequationadaptation}
\end{align}
Starting with the trained generator parameters at iteration $j=1$, Eq. \ref{optimizationequationadaptation} is solved by iteratively updating $\theta_{G}$. At the $j^{th}$ iteration, synthetic multi-coil k-space data are obtained by projecting the synthetic coil-combined image produced by the generator (i.e., $\tilde{x}^j_0 = G_{\theta_{G}^j}(\widebreve{x}_{init},0)$) through the imaging operator, where $\theta_{G}^j$ represents the parameters of the generator at $j^{th}$ iteration. The imaging operator is derived from estimated coil sensitivities $B$ \citep{Uecker2014} and the subject's prescribed sampling mask $\Omega$. After $J$ iterations, $\theta_{G}^*$ is taken as $\theta_{G}^J$, and the final reconstruction ($\widebreve{x}_{fin}$) is computed as the synthetic image produced by the generator:
\begin{align}
 \widebreve{x}_{fin}=G_{\theta_{G}^J}(\widebreve{x}_{init}, 0)
\label{eq:adadiffrec}
\end{align}

\section{Methods}

\subsection{Datasets}
Demonstrations were performed on multi-contrast brain MRI from IXI (http://brain-development.org/ixi-dataset/) and fastMRI \citep{fastmri} datasets. In IXI, coil-combined magnitude images for T\SB{1}-, T\SB{2}- and PD-weighted acquisitions were analyzed as single-coil data. Sequence parameters were repetition time (TR)=9.813ms, echo time (TE)=4.603ms, flip angle=$8^{\circ}$ for T\SB{1} scans, TR=8178ms, TE=100ms, flip angle=$90^{\circ}$ for T\SB{2} scans, TR=8178ms, TE=8ms, flip angle=$90^{\circ}$ for PD scans, and a voxel size of 0.94x0.94x1.2mm$^3$ for all. The training, validation and test sets contained (21, 15, 30) subjects resulting in (2268, 1620, 3240) cross-sections across the three contrasts. In fastMRI, multi-coil complex k-space data for T\SB{1}-, T\SB{2}- and FLAIR-weighted acquisitions were analyzed. Since MRI scans were conducted at separate sites with heterogeneous protocols, only data with at least 10 cross-sections and 20 coil elements were selected. To improve the computational efficiency of reconstruction models, geometric coil compression was used to map multi-coil data onto 5 or 10 virtual coils that preserved over 95\% and 98\% of the energy in the original data, respectively \citep{Zhang2013}. The training, validation, test sets included (240, 60, 120) subjects resulting in (2400, 600, 1200) cross-sections across contrasts. Data were retrospectively undersampled in the transverse plane (i.e., anterior-posterior and left-right dimensions) using variable-density random undersampling \citep{Lustig2007}. A normal sampling density was assumed with covariance adjusted to achieve acceleration rates of R=4x, 8x or 12x. Coil sensitivities were estimated from a fully-sampled central calibration region via ESPIRiT \citep{Uecker2014}. Volumetric k-space data undersampled in the transverse plane were inverse Fourier transformed across the fully-sampled superior/inferior dimension. Then, 2D cross-sections across the fully-sampled dimension were individually reconstructed \citep{Haldar2016,MoDl}.

\subsection{Network Architecture}

AdaDiff implemented reverse diffusion steps via an adversarial mapper comprising a generator and a discriminator (Fig. \ref{fig:architecture}). The generator followed a residual encoder-decoder structure to help project the image sample between consecutive time steps, given time index $t$ (i.e., the current time step value in the diffusion process) and a set of random normal variables $z$ \citep{DiffNvidia}. A total of 6 encoder stages were used, each stage containing 2 flat residual blocks followed by a downsampling residual block (by a factor of 2). Attention layers were used in the last two stages, and downsampling was omitted in the final stage. A total of 6 decoder stages were used, each stage containing 3 flat residual blocks followed by an upsampling residual block (by a factor of 2). An attention layer was used in the second stage, and upsampling was omitted in the final stage. Feature maps in each generator block were modulated via adaptive group normalization given a latent embedding vector \citep{adagn}. This vector was computed via a multi-layer perceptron (MLP) containing 8 fully-connected (FC) layers to embed random normal variables $z$ \citep{StyleGAN1}. The discriminator followed a residual encoder structure \citep{residual2016} to distinguish actual versus synthetic image samples. A total of 6 encoder stages were used, each containing a downsampling residual block (by a factor of 2). A final MLP with 1 FC layer was used for discrimination. All generator and discriminator blocks received a time embedding vector computed by projecting a sinusoidal encoding of the time index through a 2-layer MLP \citep{song2020score}, and this vector was added as a channel-specific bias term onto feature maps. Upsampling was performed by inserting intermediate zero-valued pixels and convolving with a learnable finite-impulse-response (FIR) filter, and downsampling was performed by convolving with a learnable FIR and discarding intermediate pixels as described in \citep{StyleGAN2}. SiLU activation functions were used in generator blocks and MLP layers, and leaky ReLU functions with negative slope 0.2 were used in discriminator blocks. Both the generator and discriminator employed 2 channels to represent the real and imaginary parts of images.

\begin{table}[t]
\centering
\captionsetup{justification=justified,singlelinecheck=false}
\caption{Within-domain performance for T\SB{1}-, T\SB{2}-, PD-weighted contrasts in IXI at R=4x-8x. PSNR (dB) and SSIM (\%) are reported as mean$\pm$std across subjects. Boldface marks the method with the highest performance metric.}
\resizebox{\columnwidth}{!}{%
\begin{tabular}{cc|c|c|c|c|c|c|c|c|}
\cline{3-9}
                                             &      & \multicolumn{1}{c|}{LORAKS} & \multicolumn{1}{c|}{rGAN} & \multicolumn{1}{c|}{MoDL} & \multicolumn{1}{c|}{GAN\SB{prior}} & \multicolumn{1}{c|}{DDPM} &
                                             \multicolumn{1}{c|}{DiffRecon} &
                                             \multicolumn{1}{c|}{AdaDiff} \\ \cline{3-9}
                                             \multicolumn{2}{c|}{ } & \multicolumn{7}{|c|}{R=4x} \\
                                             \hline
\multicolumn{1}{|c|}{\multirow{2}{*}{T\SB{1}}} & PSNR &31.2$\pm$2.2 &36.2$\pm$1.0 &41.7$\pm$1.3 &38.0$\pm$1.4 &40.8$\pm$1.3 &40.5$\pm$1.0 &{\textbf{42.1$\pm$1.6}}       \\ \cline{2-9} 
\multicolumn{1}{|c|}{}                       & SSIM &83.0$\pm$3.7 &95.0$\pm$1.0 &99.0$\pm$0.2 &96.8$\pm$1.3 &98.7$\pm$0.3 &98.5$\pm$0.3 &{\textbf{99.1$\pm$0.3}}       \\ \hline
\multicolumn{1}{|c|}{\multirow{2}{*}{T\SB{2}}} & PSNR &32.3$\pm$2.1 &35.0$\pm$0.7 &41.4$\pm$1.2 &36.6$\pm$1.8 &40.2$\pm$1.0 &37.7$\pm$1.4 &{\textbf{41.9$\pm$1.6}}       \\ \cline{2-9} 
\multicolumn{1}{|c|}{}                       & SSIM &79.5$\pm$2.9 &90.3$\pm$0.7 &98.6$\pm$0.2 &94.4$\pm$3.2 &98.1$\pm$0.2 &96.5$\pm$0.7 &{\textbf{98.9$\pm$0.2}}      \\ \hline
\multicolumn{1}{|c|}{\multirow{2}{*}{PD}}    & PSNR &31.3$\pm$2.6 &35.8$\pm$1.0 &42.0$\pm$1.5 &37.6$\pm$1.9 &40.9$\pm$1.4 &39.0$\pm$0.4 &{\textbf{42.6$\pm$1.9}}       \\ \cline{2-9} 
\multicolumn{1}{|c|}{}                       & SSIM &73.9$\pm$3.8 &91.0$\pm$0.9 &98.8$\pm$0.2 &95.8$\pm$2.1 &98.5$\pm$0.2 &97.4$\pm$0.3 &{\textbf{99.1$\pm$0.2}}       \\
\hline
                                             \multicolumn{2}{c|}{ } & \multicolumn{7}{|c|}{R=8x} \\
                                             \hline
\multicolumn{1}{|c|}{\multirow{2}{*}{T\SB{1}}} & PSNR &28.0$\pm$1.7 &32.4$\pm$1.1 &36.2$\pm$1.2 &32.8$\pm$1.3 &35.3$\pm$1.2 &36.0$\pm$0.8 &{\textbf{36.3$\pm$1.5}}       \\ \cline{2-9} 
\multicolumn{1}{|c|}{}                       & SSIM &77.4$\pm$4.5 &91.8$\pm$1.5 &97.5$\pm$0.6 &93.2$\pm$2.4 &96.8$\pm$0.7 &97.5$\pm$0.4 &{\textbf{97.6$\pm$0.6}}       \\ \hline
\multicolumn{1}{|c|}{\multirow{2}{*}{T\SB{2}}} & PSNR &28.8$\pm$1.7 &31.2$\pm$0.7 &35.9$\pm$1.2 &32.1$\pm$1.6 &34.6$\pm$1.0 &34.9$\pm$0.4 &{\textbf{36.0$\pm$1.5}}       \\ \cline{2-9} 
\multicolumn{1}{|c|}{}                       & SSIM &72.9$\pm$3.6 &85.5$\pm$1.0 &96.6$\pm$0.5 &91.1$\pm$4.0 &95.8$\pm$0.6 &95.5$\pm$0.4 &{\textbf{97.3$\pm$0.6}}      \\ \hline
\multicolumn{1}{|c|}{\multirow{2}{*}{PD}}    & PSNR &28.0$\pm$2.2 &32.0$\pm$1.1 &36.3$\pm$1.5 &32.9$\pm$1.8 &35.2$\pm$1.2 &35.4$\pm$0.5 &{\textbf{36.6$\pm$1.8}}       \\ \cline{2-9} 
\multicolumn{1}{|c|}{}                       & SSIM &66.6$\pm$4.7 &86.4$\pm$1.7 &96.7$\pm$0.7 &92.7$\pm$2.9 &96.5$\pm$0.6 &96.4$\pm$0.3 &{\textbf{97.6$\pm$0.6}}      \\ \hline

\end{tabular}%
}
\label{tab:IXI_withindomain}
\end{table}
\begin{table}[!t]
\centering
\captionsetup{justification=justified,singlelinecheck=false}
\caption{Within-domain performance for T\SB{1}-, T\SB{2}-, FLAIR- (FL.) weighted contrasts in fastMRI at R=4x-8x.}
\resizebox{\columnwidth}{!}{%
\begin{tabular}{cc|c|c|c|c|c|c|c|c|}
\cline{3-9}
                                             &      & \multicolumn{1}{c|}{LORAKS} & \multicolumn{1}{c|}{rGAN} & \multicolumn{1}{c|}{MoDL} & \multicolumn{1}{c|}{GAN\SB{prior}} & \multicolumn{1}{c|}{DDPM} &                  \multicolumn{1}{c|}{DiffRecon} &
                                             \multicolumn{1}{c|}{AdaDiff} \\ \cline{3-9}
                                             \multicolumn{2}{c|}{ } & \multicolumn{7}{|c|}{R=4x} \\
                                             \hline
\multicolumn{1}{|c|}{\multirow{2}{*}{T\SB{1}}} & PSNR  &34.0$\pm$2.3 &38.0$\pm$1.0 &39.8$\pm$1.3 &31.7$\pm$1.6 &38.2$\pm$1.7 &38.6$\pm$1.5 &{\textbf{40.2$\pm$1.7}}       \\ \cline{2-9} 
\multicolumn{1}{|c|}{}                       & SSIM  &83.8$\pm$4.8 &94.4$\pm$1.2 &95.7$\pm$1.2 &74.6$\pm$4.3 &92.5$\pm$7.6 &94.1$\pm$1.8 &{\textbf{95.9$\pm$1.4}}       \\ \hline
\multicolumn{1}{|c|}{\multirow{2}{*}{T\SB{2}}} & PSNR  &34.8$\pm$1.0 &35.3$\pm$0.7 &36.7$\pm$0.8 &30.4$\pm$0.8 &37.5$\pm$0.6 &{\textbf{39.1$\pm$0.7}} &37.7$\pm$0.8       \\ \cline{2-9}  
\multicolumn{1}{|c|}{}                       & SSIM  &91.5$\pm$1.5 &94.7$\pm$0.6 &96.0$\pm$0.5 &79.6$\pm$2.2 &95.7$\pm$0.5 &{\textbf{96.8$\pm$0.4}} &96.2$\pm$0.4      \\ \hline
\multicolumn{1}{|c|}{\multirow{2}{*}{FL}}    & PSNR   &28.6$\pm$3.2 &34.6$\pm$1.9 &36.0$\pm$2.2 &28.2$\pm$2.6 &34.1$\pm$2.9 &35.4$\pm$2.3 &{\textbf{36.2$\pm$2.6}}       \\ \cline{2-9} 
\multicolumn{1}{|c|}{}                       & SSIM  &76.8$\pm$9.6 &91.0$\pm$4.0 &{\textbf{92.7$\pm$4.0}} &72.3$\pm$8.7 &88.3$\pm$6.9 &91.7$\pm$4.7 &92.5$\pm$4.8       \\ \hline
                                             \multicolumn{2}{c|}{ } & \multicolumn{7}{|c|}{R=8x} \\
                                             \hline
\multicolumn{1}{|c|}{\multirow{2}{*}{T\SB{1}}} & PSNR  &34.1$\pm$2.0 &35.6$\pm$0.9 &37.1$\pm$1.1 &26.7$\pm$1.2 &36.2$\pm$1.3 &34.7$\pm$1.0 &{\textbf{37.2$\pm$1.5}}       \\ \cline{2-9} 
\multicolumn{1}{|c|}{}                       & SSIM   &85.3$\pm$4.5 &92.2$\pm$1.6 &{\textbf{93.5$\pm$1.6}} &57.1$\pm$4.5 &91.1$\pm$2.1 &89.4$\pm$1.8 &{\textbf{93.5$\pm$2.2}}       \\ \hline
\multicolumn{1}{|c|}{\multirow{2}{*}{T\SB{2}}} & PSNR  &33.9$\pm$0.7 &33.0$\pm$0.7 &33.9$\pm$0.8 &25.9$\pm$0.8 &34.8$\pm$0.6 &35.2$\pm$0.5 &{\textbf{35.3$\pm$0.8}}       \\ \cline{2-9} 
\multicolumn{1}{|c|}{}                       & SSIM   &92.6$\pm$1.0 &92.8$\pm$0.8 &93.9$\pm$0.7 &66.0$\pm$2.9 &94.2$\pm$0.6 &93.9$\pm$0.7 &{\textbf{94.4$\pm$0.7}}       \\ \hline
\multicolumn{1}{|c|}{\multirow{2}{*}{FL}}    & PSNR   &28.9$\pm$3.2 &33.0$\pm$1.9 &{\textbf{33.7$\pm$1.9}} &24.0$\pm$1.9 &32.8$\pm$2.2 &32.8$\pm$1.6 &{\textbf{33.7$\pm$2.4}}       \\ \cline{2-9} 
\multicolumn{1}{|c|}{}                       & SSIM   &77.5$\pm$10.2 &88.3$\pm$5.1 &{\textbf{88.9$\pm$4.8}} &57.1$\pm$7.7 &86.5$\pm$6.6 &87.3$\pm$4.8 &88.8$\pm$6.2       \\ \hline

\end{tabular}%
}
\label{tab:fastMRI_withindomain}
\end{table}

\begin{figure*}[!ht]
\vspace{-0.2cm}
\centering
\ifx\fastCompile\undefined
\includegraphics[width=1\textwidth]{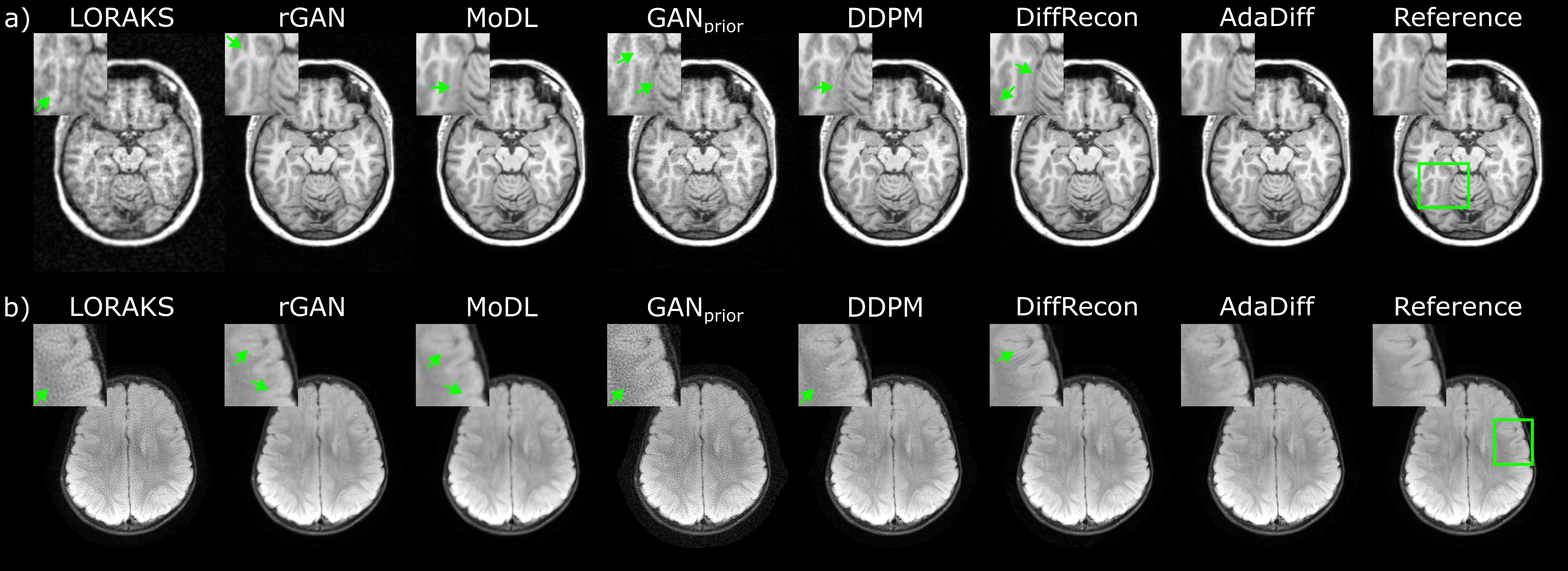}
\fi
\captionsetup{justification   = justified,singlelinecheck = false}
\caption{Within-domain reconstructions at R=4x. Results are shown for \textbf{(a)} T\SB{1}-weighted acquisitions in IXI, and \textbf{(b)} FLAIR-weighted acquisitions in fastMRI. Reconstructed images are given along with the reference image derived from fully-sampled acquisitions, and zoom-in windows and arrows are included to highlight differences among methods. LORAKS and GAN\SB{prior} show high noise amplification, rGAN shows residual aliasing, and MoDL shows visible spatial blurring despite high performance in quantitative metrics. Among diffusion models, DDPM has relatively higher noise and DiffRecon shows local ringing artifacts near tissue boundaries. AdaDiff produces high-quality reconstructions with lower artifacts/noise and clearer tissue depiction than competing methods. \vspace{-0.3cm}}
\label{fig:within_domain}
\end{figure*}

\begin{table}[t]
\centering
\captionsetup{justification=justified,singlelinecheck=false}
\caption{Cross-domain performance for T\SB{1}-, T\SB{2}-, PD-weighted contrasts in IXI. Results listed for training at R=4x, testing at R=8x and R=12x.} 
\resizebox{\columnwidth}{!}{%
\begin{tabular}{cc|c|c|c|c|c|c|c|c|}
\cline{3-9}
                                             &      & \multicolumn{1}{c|}{LORAKS} & \multicolumn{1}{c|}{rGAN} & \multicolumn{1}{c|}{MoDL} & \multicolumn{1}{c|}{GAN\SB{prior}} & \multicolumn{1}{c|}{DDPM} &                  \multicolumn{1}{c|}{DiffRecon} &
                                             \multicolumn{1}{c|}{AdaDiff} \\ \cline{3-9}
                                             \multicolumn{2}{c|}{ } & \multicolumn{7}{|c|}{R=8x} \\
                                             \hline
\multicolumn{1}{|c|}{\multirow{2}{*}{T\SB{1}}} & PSNR &28.0$\pm$1.7 &31.6$\pm$1.0 &34.7$\pm$1.3 &32.8$\pm$1.3 &35.3$\pm$1.2 &36.0$\pm$0.8 &{\textbf{36.3$\pm$1.5}}       \\ \cline{2-9} 
\multicolumn{1}{|c|}{}                       & SSIM &77.4$\pm$4.5 &90.7$\pm$1.6 &96.7$\pm$0.8 &93.2$\pm$2.4 &96.8$\pm$0.7 &97.5$\pm$0.4 &{\textbf{97.6$\pm$0.6}}       \\ \hline
\multicolumn{1}{|c|}{\multirow{2}{*}{T\SB{2}}} & PSNR &28.8$\pm$1.7 &31.1$\pm$0.8 &34.2$\pm$1.4 &32.1$\pm$1.6 &34.6$\pm$1.0 &34.9$\pm$0.4 &{\textbf{36.0$\pm$1.5}}       \\ \cline{2-9} 
\multicolumn{1}{|c|}{}                       & SSIM &72.9$\pm$3.6 &86.2$\pm$1.1 &95.4$\pm$0.8 &91.1$\pm$4.0 &95.8$\pm$0.6 &95.5$\pm$0.4 &{\textbf{97.3$\pm$0.6}}       \\ \hline
\multicolumn{1}{|c|}{\multirow{2}{*}{PD}}    & PSNR &28.0$\pm$2.2 &31.6$\pm$1.2 &34.9$\pm$1.7 &32.9$\pm$1.8 &35.2$\pm$1.2 &35.4$\pm$0.5 &{\textbf{36.6$\pm$1.8}}       \\ \cline{2-9} 
\multicolumn{1}{|c|}{}                       & SSIM &66.6$\pm$4.7 &85.6$\pm$1.9 &94.9$\pm$0.8 &92.7$\pm$2.9 &96.5$\pm$0.6 &96.4$\pm$0.3 &{\textbf{97.6$\pm$0.6}}       \\ \hline
                                             \multicolumn{2}{c|}{ } & \multicolumn{7}{|c|}{R=12x} \\
                                             \hline
\multicolumn{1}{|c|}{\multirow{2}{*}{T\SB{1}}} & PSNR  &26.5$\pm$1.6 &29.2$\pm$1.0 &31.6$\pm$1.3 &31.1$\pm$1.3 &32.7$\pm$1.1 &32.9$\pm$0.9 &{\textbf{33.4$\pm$1.3}}       \\ \cline{2-9} 
\multicolumn{1}{|c|}{}                       & SSIM   &73.7$\pm$5.3 &86.7$\pm$2.4 &92.7$\pm$1.5 &92.1$\pm$1.9 &95.2$\pm$1.0 &95.3$\pm$0.8 &{\textbf{96.2$\pm$1.0}}       \\ \hline
\multicolumn{1}{|c|}{\multirow{2}{*}{T\SB{2}}} & PSNR  &27.3$\pm$1.6 &28.9$\pm$0.9 &31.1$\pm$1.3 &29.5$\pm$1.5 &32.0$\pm$1.0 &31.9$\pm$0.8 &{\textbf{33.1$\pm$1.5}}       \\ \cline{2-9} 
\multicolumn{1}{|c|}{}                       & SSIM   &69.2$\pm$4.1 &81.4$\pm$1.7 &89.2$\pm$1.6 &86.3$\pm$4.8 &94.0$\pm$1.0 &93.1$\pm$0.7 &{\textbf{95.7$\pm$1.0}}       \\ \hline
\multicolumn{1}{|c|}{\multirow{2}{*}{PD}}    & PSNR   &26.4$\pm$2.0 &29.3$\pm$1.2 &31.7$\pm$1.6 &31.1$\pm$1.6 &32.6$\pm$1.2 &32.7$\pm$1.0 &{\textbf{33.9$\pm$1.8}}       \\ \cline{2-9} 
\multicolumn{1}{|c|}{}                       & SSIM   &62.1$\pm$5.2 &80.2$\pm$2.3 &85.5$\pm$2.4 &90.4$\pm$3.3 &94.8$\pm$1.0 &94.5$\pm$0.7 &{\textbf{96.2$\pm$1.0}}       \\ \hline
\vspace{-0.75cm}
\end{tabular}%
}
\label{tab:IXI_crossdomain}
\end{table}

\begin{table}[t]
\centering
\captionsetup{justification=justified,singlelinecheck=false}
\caption{Cross-domain performance for T\SB{1}-, T\SB{2}-, and FLAIR- (FL.) weighted contrasts in fastMRI. Results for training at R=4x, testing at R=8x and R=12x.} 
\resizebox{\columnwidth}{!}{%
\begin{tabular}{cc|c|c|c|c|c|c|c|c|}
\cline{3-9}
                                             &      & \multicolumn{1}{c|}{LORAKS} & \multicolumn{1}{c|}{rGAN} & \multicolumn{1}{c|}{MoDL} & \multicolumn{1}{c|}{GAN\SB{prior}} & \multicolumn{1}{c|}{DDPM} &                  \multicolumn{1}{c|}{DiffRecon} &
                                             \multicolumn{1}{c|}{AdaDiff} \\ \cline{3-9}
                                             \multicolumn{2}{c|}{ } & \multicolumn{7}{|c|}{R=8x} \\
                                             \hline
\multicolumn{1}{|c|}{\multirow{2}{*}{T\SB{1}}} & PSNR &34.1$\pm$2.0 &36.0$\pm$0.9 &36.4$\pm$1.0 &26.7$\pm$1.2 &36.2$\pm$1.3 &34.7$\pm$1.0 &{\textbf{37.2$\pm$1.5}}       \\ \cline{2-9} 
\multicolumn{1}{|c|}{}                       & SSIM  &85.3$\pm$4.5 &92.4$\pm$1.5 &93.3$\pm$1.6 &57.1$\pm$4.5 &91.1$\pm$2.1 &89.4$\pm$1.8 &{\textbf{93.5$\pm$2.2}}       \\ \hline
\multicolumn{1}{|c|}{\multirow{2}{*}{T\SB{2}}} & PSNR &33.9$\pm$0.7 &33.0$\pm$0.7 &32.9$\pm$0.8 &25.9$\pm$0.8 &34.8$\pm$0.6 &35.2$\pm$0.5 &{\textbf{35.3$\pm$0.8}}       \\ \cline{2-9} 
\multicolumn{1}{|c|}{}                       & SSIM &92.6$\pm$1.0 &92.2$\pm$0.7 &93.4$\pm$0.8 &66.0$\pm$2.9 &94.2$\pm$0.6 &93.9$\pm$0.7  &{\textbf{94.4$\pm$0.6}}       \\ \hline
\multicolumn{1}{|c|}{\multirow{2}{*}{FL}} & PSNR &28.9$\pm$3.2 &32.7$\pm$1.8 &33.2$\pm$1.9 &24.0$\pm$1.9 &32.8$\pm$2.2 &32.8$\pm$1.6  &{\textbf{33.7$\pm$2.4}}       \\ \cline{2-9} 
\multicolumn{1}{|c|}{}                       & SSIM &77.5$\pm$10.2 &87.7$\pm$4.9 &{\textbf{88.8$\pm$4.8}} &57.1$\pm$7.7 &86.5$\pm$6.6 &87.3$\pm$4.8 &{\textbf{88.8$\pm$6.2}}       \\ \hline
                                             \multicolumn{2}{c|}{ } & \multicolumn{7}{|c|}{R=12x} \\
                                             \hline
\multicolumn{1}{|c|}{\multirow{2}{*}{T\SB{1}}} & PSNR  &34.1$\pm$1.6 &34.6$\pm$0.9 &34.7$\pm$1.0 &24.4$\pm$1.1 &35.9$\pm$0.9 &33.0$\pm$0.8 &{\textbf{35.2$\pm$1.6}}       \\ \cline{2-9} 
\multicolumn{1}{|c|}{}                       & SSIM   &86.9$\pm$4.0 &91.1$\pm$1.6 &{\textbf{91.7$\pm$1.8}} &48.8$\pm$5.0 &91.2$\pm$1.5 &86.4$\pm$1.9 &91.2$\pm$2.7       \\ \hline
\multicolumn{1}{|c|}{\multirow{2}{*}{T\SB{2}}} & PSNR  &32.9$\pm$0.7 &31.5$\pm$0.8 &31.2$\pm$0.8 &23.0$\pm$0.8 &33.2$\pm$0.5 &33.2$\pm$0.5 &{\textbf{33.9$\pm$0.8}}       \\ \cline{2-9} 
\multicolumn{1}{|c|}{}                       & SSIM   &92.2$\pm$0.9 &90.5$\pm$1.0 &91.4$\pm$1.0 &56.1$\pm$3.4 &92.8$\pm$0.8 &91.3$\pm$1.0 &{\textbf{93.2$\pm$0.8}}       \\ \hline
\multicolumn{1}{|c|}{\multirow{2}{*}{FL}}    & PSNR   &29.5$\pm$3.0 &31.6$\pm$1.7 &31.8$\pm$1.7 &22.1$\pm$1.6 &32.2$\pm$1.7 &31.4$\pm$1.4 &{\textbf{32.3$\pm$2.3}}       \\ \cline{2-9} 
\multicolumn{1}{|c|}{}                       & SSIM   &79.2$\pm$9.9 &85.5$\pm$5.3 &86.2$\pm$5.3 &50.2$\pm$7.3 &86.3$\pm$5.7 &84.1$\pm$4.8 &{\textbf{86.4$\pm$6.9}}       \\ \hline
\vspace{-0.75cm}
\end{tabular}%
}
\label{tab:fastMRI_crossdomainAcceleration}
\end{table}

\subsection{Competing Methods}
AdaDiff was demonstrated for MRI reconstruction against a traditional method (LORAKS), conditional models (rGAN, MoDL), and unconditional models (GAN\SB{prior}, DDPM, DiffRecon). Conditional models were trained to map inverse Fourier transform of undersampled data onto ground-truth images derived from fully-sampled data. Unconditional models were trained to generate coil-combined MR image samples derived from fully-sampled data. An exponentially decreasing noise scheduler with parameters $\beta_{min}, \beta_{max}$=$\{0.1, 20 \}$ was adopted for all diffusion models \citep{song2020score}. Hyperparameter selection was performed via one-fold cross-validation. Deep models were trained via the Adam optimizer using the decay rates $\beta_1$=0.5 and $\beta_2$=0.9. Prior adaptation during inference was performed via the Adam optimizer at $\beta_1$=0.5 and $\beta_2$=0.9. All deep models were executed on Nvidia RTX 3090 GPUs via PyTorch.  

\textbf{AdaDiff}: Hyperparameters for AdaDiff were 6x10$^{-3}$ learning rate, 500 epochs, $k$=125 step size, $T/k$=8 diffusion steps for training; 8 iterations combining a reverse diffusion step and a data-consistency projection for rapid diffusion; and 10$^{-3}$ learning rate, 1000 iterations for prior adaptation.

\textbf{LORAKS}: An autocalibrated low-rank reconstruction was implemented \citep{Haldar2016}. The k-space neighborhood radius and the rank of the system matrix were selected as: (2,6) for IXI, and (2,30) for fastMRI \citep{elmas2022federated}.

\textbf{rGAN}: A conditional GAN model was implemented with architecture and loss functions in \citep{rgan}. Hyperparameters were selected as 2x10$^{-4}$ learning rate, 100 epochs, adversarial and pixel-wise loss weights of (1,100) for training \citep{rgan}.

\textbf{MoDL}: A conditional unrolled model that interleaves data-consistency blocks with convolutional layers was implemented \citep{MoDl}. The architecture and loss function were adopted from \cite{Dar2021ismrm}. Hyperparameters were selected as 10$^{-3}$ learning rate, 200 epochs for training \citep{Dar2021ismrm}.

\textbf{GAN\SB{prior}}: An unconditional GAN model that performs prior adaptation was implemented \citep{Knoll2019inverseGANs}. The architecture and loss function were adopted from \citep{StyleGAN1}. GAN\SB{prior} performed prior adaptation by minimizing a data-consistency loss as in AdaDiff. Hyperparameters were selected as 10$^{-3}$ learning rate, 3000 epochs for training; 10$^{-2}$ learning rate, 1000 iterations for inference \citep{elmas2022federated}.

\textbf{DDPM}: An unconditional diffusion model was implemented with architecture and loss functions in \citep{DDPM}. Hyperparameters were selected as 10$^{-4}$ learning rate, $k$=1 step size, $T/k$=1000 diffusion steps, 65 epochs for training; 1000 iterations combining a reverse diffusion step and a data-consistency projection for inference \citep{DDPM}.

\textbf{DiffRecon}: An unconditional diffusion model was implemented with architecture and loss functions in \citep{peng2022}. Hyperparameters were selected as 10$^{-4}$ learning rate, $k$=1 step size, $T/k$=4000 diffusion steps, 300 epochs for training; 400 coarse and 20 fine iterations of reverse diffusion and data-consistency projection for inference \citep{peng2022}.

\subsection{Analyses}
Retrospectively undersampled acquisitions in IXI and fastMRI were reconstructed. Here, a single unified model was trained on aggregate data from multiple distinct contrasts to improve practicality given the pervasiveness of multi-contrast protocols. In each dataset, model training was accordingly performed on data pooled across multiple contrasts: (T\SB{1},T\SB{2},PD) in IXI and (T\SB{1},T\SB{2},FLAIR) in fastMRI. Training samples were randomly drawn from the pooled data, and the model was not informed regarding the contrast of the samples. Conditional models receive undersampled data as input so they are informed regarding the acceleration rate during training. We observed that training a unified model with undersampled data from mixed R values did not have a notable influence on performance. Thus, separate models were trained at each individual R value to prevent biases in performance assessments under domain shifts in R between training and test sets. Reconstruction quality was measured via peak signal-to-noise ratio (PSNR) and structural similarity (SSIM) metrics between the recovered and ground-truth images derived from fully-sampled acquisitions. In ablation studies, image fidelity was additionally characterized via Frechet inception distance (FID; \cite{FID_NIPS2017_8a1d6947}) and learned perceptual image patch similarity (LPIPS; \cite{lpips_Zhang_2018_CVPR}) metrics. Images were normalized to unity mean prior to measurements. Significance of differences in PSNR, SSIM and LPIPS between models were evaluated with non-parametric signed-rank tests. Note that FID quantifies the overall similarity of distributions across the examined set of samples as a single metric value, so significance testing was not conducted for FID.

\begin{figure*}[t]
\vspace{-0.2cm}
\centering
\ifx\fastCompile\undefined
\includegraphics[width=1\textwidth]{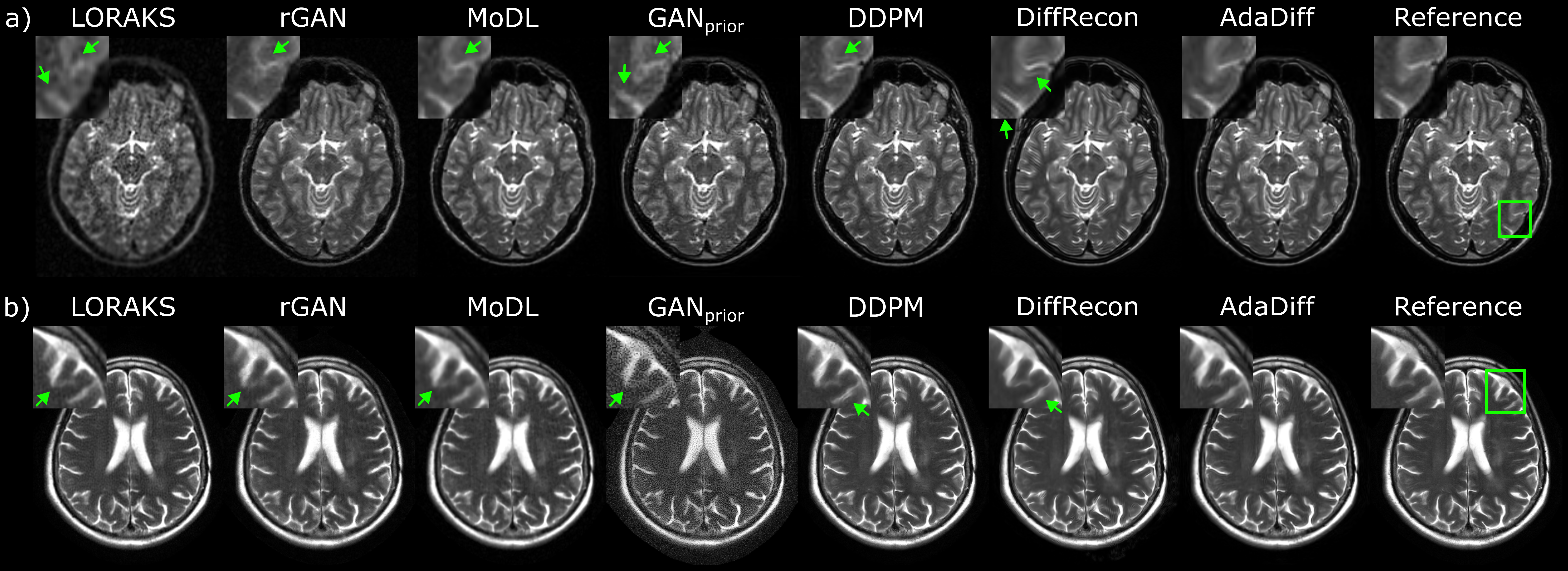}
\fi
\captionsetup{justification   = justified,singlelinecheck = false}
\caption{Cross-domain reconstructions under domain shifts in the acceleration rate. Results are shown for \textbf{(a)} T\SB{2}-weighted acquisitions at R=8x in IXI, and \textbf{(b)} T\SB{2}-weighted acquisitions at R=12x in fastMRI. Reconstructed images are given along with the reference image derived from fully-sampled acquisitions, and zoom-in windows and arrows are included to highlight differences among methods. Conditional models were trained at R=4x. LORAKS and GAN\SB{prior} show high noise amplification, rGAN and MoDL show some residual reconstruction artifacts and spatial blurring. Among diffusion models, DDPM has relatively high noise and DiffRecon has local ringing artifacts. AdaDiff reconstructs images with low artifacts/noise and a closer appearance to the reference images. 
\vspace{-0.25cm}}
\label{fig:cross_domain}
\end{figure*}

\section{Results}
\subsection{Within-Domain Reconstruction}
AdaDiff was first demonstrated for within-domain reconstructions where the imaging operator and the MR image distribution were matched between the training and test sets (e.g., trained and tested for R=4x in fastMRI). Comparisons were performed against a traditional method (LORAKS), conditional models (rGAN, MoDL), an unconditional GAN that performs prior adaptation (GAN\SB{prior}), and unconditional diffusion models that use static priors (DDPM, DiffRecon). PSNR and SSIM for competing methods are listed in Table \ref{tab:IXI_withindomain} for IXI, and in Table \ref{tab:fastMRI_withindomain} for fastMRI. In IXI, AdaDiff achieves the highest performance among competing methods across contrasts and acceleration rates (p$<$0.05), except for MoDL that performs similarly on T\SB{1} in general and on T\SB{2} at R=8x in PSNR. In fastMRI, AdaDiff again outperforms competing methods across contrasts and acceleration rates (p$<$0.05), except for MoDL that performs similarly on T\SB{1}, FLAIR in general, and DiffRecon that yields higher performance on T\SB{2} at R=4x and similar PSNR on T\SB{2} at R=8x. On average, AdaDiff outperforms the traditional method by 6.8dB PSNR and 15.8\% SSIM, conditional models by 2.0dB PSNR and 2.5\% SSIM, the adaptive GAN by 6.6dB PSNR and 15.0\% SSIM, and static diffusion models by 1.3dB PSNR and 1.4\% SSIM. These results indicate that the adaptive diffusion prior in AdaDiff helps improve reconstruction quality over both an adaptive GAN prior and static diffusion priors. Representative reconstructions are displayed in Fig. \ref{fig:within_domain}. LORAKS and GAN\SB{prior} show high noise amplification. While rGAN and MoDL have relatively low noise levels, rGAN shows residual reconstruction artifacts and MoDL suffers from spatial blurring, which can be attributed to its pixel-wise loss function. Among diffusion models, DDPM has relatively high noise whereas DiffRecon is effective in noise suppression via repeated averaging of multiple diffusion samples. DiffRecon tends to produce sharper images than AdaDiff likely due to its fine iteration steps, but this refinement can also introduce ringing artifacts near tissue boundaries by emphasizing high spatial frequencies. In contrast, AdaDiff adapts its diffusion prior to better conform to the distribution of test data, enabling it to produce high-quality reconstructions that clearly depict tissues with lower artifacts and noise than competing methods.

\begin{figure*}[t]
\centering
\ifx\fastCompile\undefined
\includegraphics[width=1\textwidth]{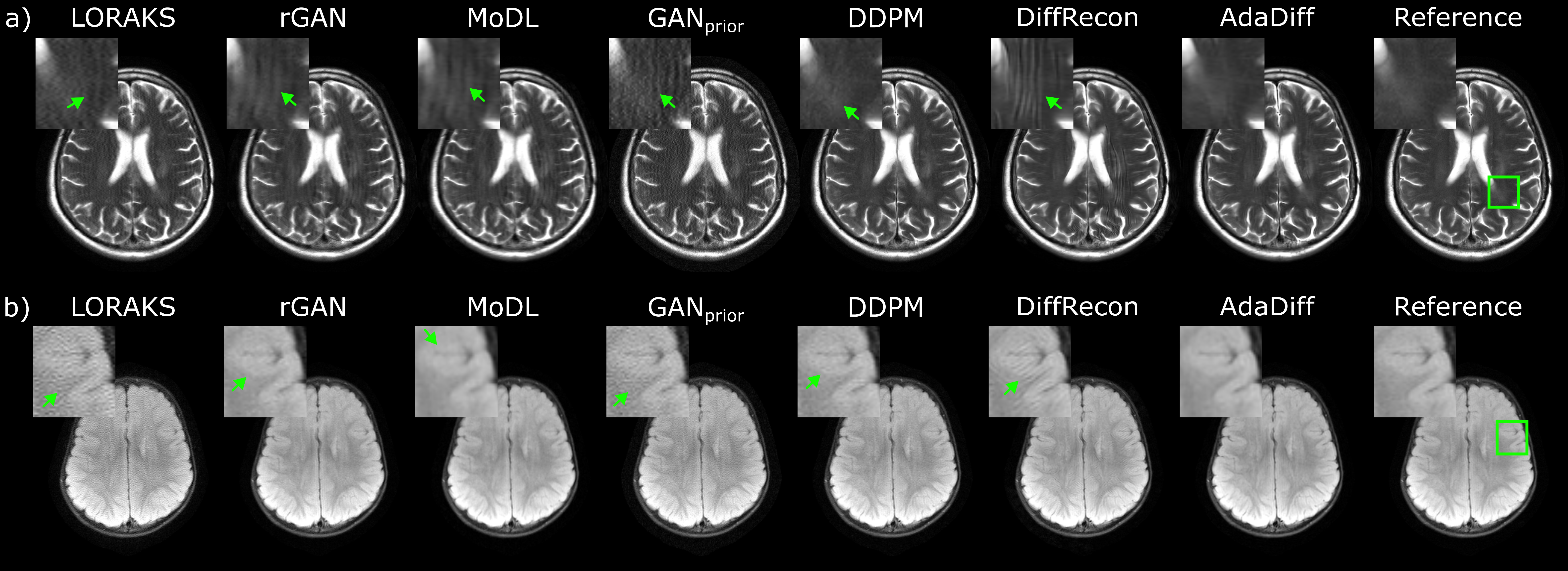}
\fi
\captionsetup{justification   = justified,singlelinecheck = false}
\caption{Cross-domain reconstructions at R=4x under domain shifts in the sampling trajectory and number of coils. Results are shown for \textbf{(a)} T\SB{2}-weighted acquisitions with 1D undersampling, and \textbf{(b)} PD-weighted acquisitions with 10 virtual coils in fastMRI. Reconstructed images are given along with the reference image derived from fully-sampled acquisitions, and zoom-in windows and arrows are included to highlight differences among methods. Conditional models were trained for 2D undersampling and 5 virtual coils. LORAKS and GAN\SB{prior} show high noise amplification, rGAN and MoDL show residual reconstruction artifacts and blurring. DDPM has relatively high noise and DiffRecon has local ringing artifacts. AdaDiff reconstructs images with low artifacts/noise.}
\label{fig:1dgaussianand10coil}
\end{figure*}

\subsection{Domain Shifts in the Imaging Operator}
We then demonstrated performance in cross-domain reconstructions where the MR image distribution was matched, albeit the imaging operator was mismatched between the training and test sets. To this end, several studies were conducted to examine the influence of varying operator attributes on reconstruction performance. First, we assessed the influence of acceleration rate by training conditional models at R=4x while testing all models at R=8x and R=12x. Note that unconditional models and LORAKS were not informed about undersampling during training. PSNR and SSIM for competing methods are listed in Table \ref{tab:IXI_crossdomain} for IXI, and in Table \ref{tab:fastMRI_crossdomainAcceleration} for fastMRI. In IXI, AdaDiff achieves the highest performance across contrasts and acceleration rates (p$<$0.05), except for DiffRecon that yields similar SSIM on T\SB{1} at R=8x. In fastMRI, AdaDiff again outperforms competing methods across contrasts and acceleration rates (p$<$0.05), except for MoDL that yields similar SSIM on FLAIR at R=8x and higher SSIM on T\SB{1} at R=12x, and DiffRecon that yields similar PSNR on T\SB{2} at R=8x. On average, AdaDiff outperforms the traditional method by 4.9dB PSNR and 16.0\% SSIM, conditional models by 2.3dB PSNR and 4.4\% SSIM, the adaptive GAN by 6.8dB PSNR and 20.6\% SSIM, and static diffusion models by 0.9dB PSNR and 1.5\% SSIM. Note that, at R=8x, the performance benefit of AdaDiff over MoDL is 1.5dB PSNR, 2.3\% SSIM under domain shift in acceleration rate, versus 0.3dB PSNR, 0.4\% SSIM in within-domain reconstruction. This difference suggests that AdaDiff's unconditional prior is more reliable against variations in acceleration rate compared to MoDL's conditional prior. Representative reconstructions are displayed in Fig. \ref{fig:cross_domain}. LORAKS and GAN\SB{prior} suffer from noise amplification, rGAN and MoDL shows residual aliasing artifacts and blurring. Similar to the within-domain case, DDPM has relatively high noise levels and DiffRecon shows local ringing artifacts in comparison to AdaDiff that maintains the closest appearance to the reference images. 

Next, we assessed the influence of domain shifts in sampling trajectory and number of coils on reconstruction performance. Under fixed acceleration rate and number of coils, sampling trajectory was varied by training conditional models based on 2D undersampling patterns while testing all models on 1D undersampling patterns. Under fixed acceleration rate and sampling trajectory, number of coils was varied by training conditional models based on 5 virtual coils and testing all models on 10 virtual coils. PSNR and SSIM for competing methods are listed in Table \ref{tab:fastMRI_AllImagingOp} for both assessments. When the sampling trajectory is varied, AdaDiff achieves the highest performance among competing methods across tissue contrasts (p$<$0.05), except for MoDL that yields higher SSIM on FLAIR, and DDPM that yields similar PSNR on FLAIR. When the number of coils is varied, AdaDiff yields higher performance across tissue contrasts (p$<$0.05), except for MoDL that yields similar SSIM on FLAIR, and DiffRecon that yields higher performance on T\SB{2}. On average, AdaDiff outperforms the traditional method by 5.0dB PSNR and 10.0\% SSIM, conditional models by 1.6dB PSNR and 1.2\% SSIM, the adaptive GAN by 7.2dB PSNR and 20.2\% SSIM, and static diffusion models by 1.0dB PSNR and 2.0\% SSIM. Note that, at R=4x in fastMRI, the performance benefit of AdaDiff over MoDL is 1.6dB PSNR, 1.2\% SSIM under domain shift in sampling trajectory, 0.6dB PSNR, 0.2\% SSIM under domain shift in number of coils, versus 0.5dB PSNR, 0.1\% SSIM in within-domain reconstruction. These findings suggest that AdaDiff is notably more reliable than MoDL against shifts in the sampling trajectory, whereas it is similarly affected by shifts in the number of coils. Representative reconstructions are displayed in Fig. \ref{fig:1dgaussianand10coil}. LORAKS and GAN\SB{prior} suffer from noise amplification, rGAN and MoDL show residual aliasing artifacts and blurring. Similar to the within-domain case, DDPM has relatively high noise levels and DiffRecon shows local ringing artifacts in comparison to AdaDiff that maintains the closest appearance to the reference images.

\begin{table}[t]
\centering
\captionsetup{justification=justified,singlelinecheck=false}
\caption{Cross-domain performance for T\SB{1}-, T\SB{2}-, and FL.-weighted contrasts in fastMRI at R=4x. Training under 2D undersampling with 5 coils, testing under 1D undersampling with 5 coils and 2D undersampling with 10 coils.}
\resizebox{\columnwidth}{!}{%
\begin{tabular}{cc|c|c|c|c|c|c|c|c|}
\cline{3-9}
                                             &      & \multicolumn{1}{c|}{LORAKS} & \multicolumn{1}{c|}{rGAN} & \multicolumn{1}{c|}{MoDL} & \multicolumn{1}{c|}{GAN\SB{prior}} & \multicolumn{1}{c|}{DDPM} &
                                             \multicolumn{1}{c|}{DiffRecon} &
                                             \multicolumn{1}{c|}{AdaDiff} \\ \cline{3-9}
                                             \multicolumn{2}{c|}{ } & \multicolumn{7}{|c|}{1D, 5 coils} \\
                                             \hline
\multicolumn{1}{|c|}{\multirow{2}{*}{T\SB{1}}} & PSNR &33.4$\pm$1.6 &34.6$\pm$1.0 &35.3$\pm$1.1 &28.7$\pm$1.3 &35.7$\pm$1.1 &32.1$\pm$0.9 &{\textbf{36.5$\pm$1.8}}       \\ \cline{2-9} 
\multicolumn{1}{|c|}{}                       & SSIM &84.9$\pm$4.2 &91.7$\pm$1.3 &92.8$\pm$1.3 &66.8$\pm$4.2 &91.2$\pm$1.3 &85.1$\pm$2.4 &{\textbf{93.3$\pm$2.0}}       \\ \hline
\multicolumn{1}{|c|}{\multirow{2}{*}{T\SB{2}}} & PSNR &32.2$\pm$0.8 &30.7$\pm$0.8 &31.1$\pm$0.9 &28.4$\pm$0.7 &33.5$\pm$0.6 &33.0$\pm$0.5 &{\textbf{34.0$\pm$0.8}}       \\ \cline{2-9} 
\multicolumn{1}{|c|}{}                       & SSIM &91.9$\pm$1.5 &90.2$\pm$1.2 &90.9$\pm$1.2 &77.1$\pm$2.5 &93.4$\pm$0.8 &92.7$\pm$1.0 &{\textbf{94.3$\pm$0.8}}       \\ \hline
\multicolumn{1}{|c|}{\multirow{2}{*}{FL}}    & PSNR &28.1$\pm$2.5 &31.7$\pm$1.5 &32.3$\pm$1.5 &26.1$\pm$2.0 &32.7$\pm$1.6 &31.5$\pm$1.3 &{\textbf{32.8$\pm$2.0}}       \\ \cline{2-9} 
\multicolumn{1}{|c|}{}                       & SSIM &78.2$\pm$6.3 &88.3$\pm$2.7 &{\textbf{89.7$\pm$2.5}} &66.6$\pm$6.5 &88.6$\pm$3.5 &86.5$\pm$3.6 &89.3$\pm$4.3       \\ \hline
                                             \multicolumn{2}{c|}{ } & \multicolumn{7}{|c|}{2D, 10 coils} \\
                                             \hline
\multicolumn{1}{|c|}{\multirow{2}{*}{T\SB{1}}} & PSNR &33.2$\pm$1.9 &38.1$\pm$1.0 &39.7$\pm$1.3 &32.1$\pm$1.7 &38.4$\pm$1.5 &38.8$\pm$1.5 &{\textbf{40.2$\pm$1.7}}       \\ \cline{2-9} 
\multicolumn{1}{|c|}{}                       & SSIM &82.3$\pm$4.2 &94.6$\pm$1.3 &95.7$\pm$1.3 &76.4$\pm$4.4 &93.3$\pm$1.8 &94.5$\pm$1.8 &{\textbf{96.1$\pm$1.5}}       \\ \hline
\multicolumn{1}{|c|}{\multirow{2}{*}{T\SB{2}}} & PSNR &33.6$\pm$1.0 &35.4$\pm$0.7 &36.7$\pm$0.7 &30.5$\pm$0.7 &37.4$\pm$0.6 &{\textbf{39.2$\pm$0.7}} &37.7$\pm$0.8       \\ \cline{2-9} 
\multicolumn{1}{|c|}{}                       & SSIM &90.3$\pm$1.8 &95.0$\pm$0.5 &96.1$\pm$0.5 &80.6$\pm$2.0 &95.9$\pm$0.4 &{\textbf{97.1$\pm$0.3}} &96.4$\pm$0.4       \\ \hline
\multicolumn{1}{|c|}{\multirow{2}{*}{FL}}    & PSNR &27.2$\pm$3.0 &34.6$\pm$2.3 &36.0$\pm$2.5 &28.4$\pm$3.0 &34.6$\pm$2.9 &35.5$\pm$2.6 &{\textbf{36.3$\pm$3.0}}       \\ \cline{2-9} 
\multicolumn{1}{|c|}{}                       & SSIM &74.5$\pm$9.5 &91.1$\pm$5.3 &{\textbf{92.5$\pm$5.3}} &73.1$\pm$10.2 &89.6$\pm$7.5 &91.6$\pm$6.0 &92.4$\pm$6.4       \\ \hline
\end{tabular}%
}
\label{tab:fastMRI_AllImagingOp}
\end{table}

\begin{figure*}[t]
\centering
\ifx\fastCompile\undefined
\includegraphics[width=1\textwidth]{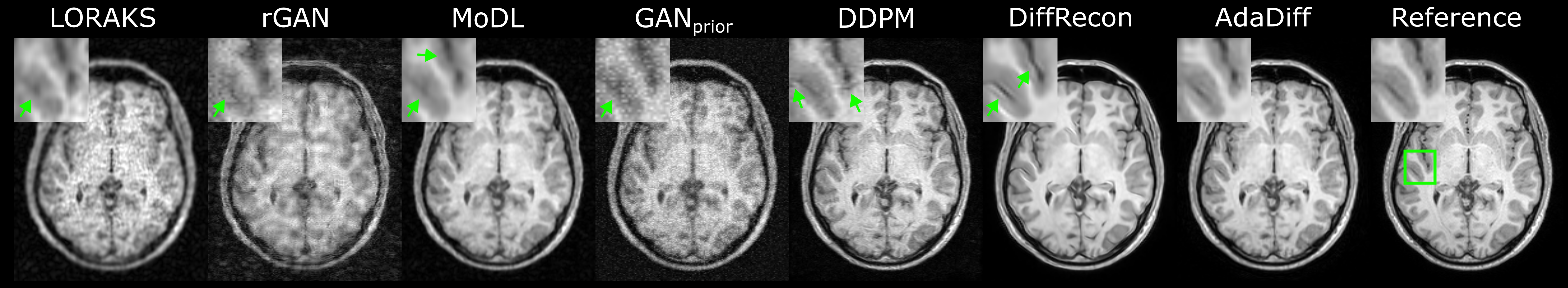}
\fi
\captionsetup{justification   = justified,singlelinecheck = false}
\caption{Cross-domain reconstructions under domain shifts in the MR image distribution. Results are shown for T\SB{1}-weighted acquisitions at R=8x. All models were trained on fastMRI and tested on IXI. The imaging operator matched between training and testing for conditional models. Reconstructed images are given along with the reference image derived from fully-sampled acquisitions, and zoom-in windows and arrows are included to highlight differences among methods. High noise amplification in LORAKS and GAN\SB{prior}, substantial residual artifacts in rGAN, and spatial blurring in MoDL are observed. While DDPM shows elevated noise and DiffRecon yields ringing artifacts, AdaDiff achieves high-fidelity reconstructions with clear tissue depiction.}
\label{fig:all_cross_domain}
\end{figure*}

\begin{table}[t]
\centering
\captionsetup{justification=justified,singlelinecheck=false}
\caption{Cross-domain performance for T\SB{1}-, T\SB{2}-, PD-weighted contrasts at R=4x-8x. Results listed for training on fastMRI, testing on IXI.}
\resizebox{\columnwidth}{!}{%
\begin{tabular}{cc|c|c|c|c|c|c|c|c|}
\cline{3-9}
                                             &      & \multicolumn{1}{c|}{LORAKS} & \multicolumn{1}{c|}{rGAN} & \multicolumn{1}{c|}{MoDL} & \multicolumn{1}{c|}{GAN\SB{prior}} & \multicolumn{1}{c|}{DDPM} &
                                             \multicolumn{1}{c|}{DiffRecon} &
                                             \multicolumn{1}{c|}{AdaDiff} \\ \cline{3-9}
                                             \multicolumn{2}{c|}{ } & \multicolumn{7}{|c|}{R=4x} \\
                                             \hline
\multicolumn{1}{|c|}{\multirow{2}{*}{T\SB{1}}} & PSNR    &31.8$\pm$3.4 &30.3$\pm$2.1 &34.1$\pm$1.8 &28.2$\pm$1.8 &35.3$\pm$1.7 &38.8$\pm$0.9 &{\textbf{41.0$\pm$2.1}}       \\ \cline{2-9} 
\multicolumn{1}{|c|}{}                       & SSIM    &83.5$\pm$5.4 &79.4$\pm$5.4 &89.3$\pm$2.4 &73.3$\pm$5.3 &90.1$\pm$2.3 &{\textbf{98.6$\pm$0.2}} &{\textbf{98.6$\pm$0.5}}       \\ \hline
\multicolumn{1}{|c|}{\multirow{2}{*}{T\SB{2}}} & PSNR    &32.6$\pm$3.6 &30.1$\pm$1.3 &32.8$\pm$1.7 &27.2$\pm$1.7 &35.4$\pm$0.8 &38.6$\pm$0.5 &{\textbf{40.5$\pm$1.9}}       \\ \cline{2-9} 
\multicolumn{1}{|c|}{}                       & SSIM    &80.8$\pm$6.4 &72.9$\pm$3.6 &81.3$\pm$3.0 &59.3$\pm$6.1 &87.8$\pm$1.1 &{\textbf{98.3$\pm$0.2}} &98.1$\pm$0.5       \\ \hline
\multicolumn{1}{|c|}{\multirow{2}{*}{PD}}    & PSNR   &32.0$\pm$3.9 &29.6$\pm$1.9 &32.9$\pm$2.0 &28.4$\pm$1.9 &34.5$\pm$0.9 &39.1$\pm$0.7 &{\textbf{40.8$\pm$2.0}}      \\ \cline{2-9} 
\multicolumn{1}{|c|}{}                       & SSIM   &75.2$\pm$7.9 &68.7$\pm$5.3 &79.3$\pm$4.3 &64.4$\pm$6.2 &83.6$\pm$1.7 &{\textbf{98.7$\pm$0.2}} &98.0$\pm$0.6       \\
\hline
                                             \multicolumn{2}{c|}{ } & \multicolumn{7}{|c|}{R=8x} \\
                                             \hline
\multicolumn{1}{|c|}{\multirow{2}{*}{T\SB{1}}} & PSNR    &28.4$\pm$2.6 &27.0$\pm$1.9 &30.8$\pm$1.6 &22.9$\pm$1.1 &32.1$\pm$1.7 &34.4$\pm$0.8 &{\textbf{35.6$\pm$2.0}}       \\ \cline{2-9} 
\multicolumn{1}{|c|}{}                       & SSIM   &78.0$\pm$6.6 &71.1$\pm$6.9 &85.5$\pm$3.1 &55.9$\pm$6.4 &86.5$\pm$2.9 &{\textbf{96.9$\pm$0.5}} &96.1$\pm$1.3       \\ \hline
\multicolumn{1}{|c|}{\multirow{2}{*}{T\SB{2}}} & PSNR   &29.1$\pm$2.8 &27.4$\pm$0.9 &30.0$\pm$1.5 &22.8$\pm$1.4 &32.0$\pm$0.9 &34.2$\pm$0.4 &{\textbf{35.1$\pm$1.9}}       \\ \cline{2-9} 
\multicolumn{1}{|c|}{}                       & SSIM   &74.5$\pm$7.8 &66.1$\pm$3.2 &76.9$\pm$3.3 &41.5$\pm$7.1 &82.9$\pm$1.6 &{\textbf{96.5$\pm$0.5}} &95.8$\pm$1.3       \\ \hline
\multicolumn{1}{|c|}{\multirow{2}{*}{PD}}    & PSNR   &28.6$\pm$3.2 &26.1$\pm$1.5 &30.0$\pm$1.8 &22.7$\pm$1.8 &31.4$\pm$1.0 &34.5$\pm$0.6 &{\textbf{35.6$\pm$2.0}}      \\ \cline{2-9} 
\multicolumn{1}{|c|}{}                       & SSIM   &68.0$\pm$9.4 &58.1$\pm$5.4 &73.9$\pm$4.6 &43.1$\pm$8.1 &78.6$\pm$2.1 &{\textbf{96.9$\pm$0.5}} &95.2$\pm$1.6       \\ \hline

\end{tabular}%
}
\label{tab:all_crossdomain}
\end{table}

\subsection{Domain Shifts in the Image Distribution}
We also examined cross-domain reconstruction where the imaging operator was matched, albeit the MR image distribution was mismatched between the training and test sets. For this purpose, training was performed on fastMRI and testing was performed on IXI. PSNR and SSIM for competing methods are listed in Table \ref{tab:all_crossdomain}. In general, AdaDiff achieves the highest performance among competing methods across tissue contrasts and acceleration rates (p$<$0.05), except for DiffRecon that yields modestly higher SSIM. On average, AdaDiff outperforms the traditional method by 7.7dB PSNR and 20.3\% SSIM, conditional models by 8.0dB PSNR and 21.8\% SSIM, the adaptive GAN by 12.7dB PSNR and 40.7\% SSIM, and static diffusion models by 3.1dB PSNR and 5.7\% SSIM. Note that, compared to domain shifts in the imaging operator, a domain shift in the image distribution induces more notable performance losses for competing methods including static diffusion models. Although the adaptive GAN model uses prior adaptation, its relatively poor performance can be attributed to the low representational diversity of adversarial priors. In contrast, the adaptive diffusion prior in AdaDiff maintains high reconstruction performance. Representative reconstructions are displayed in Fig. \ref{fig:all_cross_domain}. High noise amplification in LORAKS and GAN\SB{prior}, substantial residual artifacts in rGAN, and spatial blurring in MoDL are observed. While DDPM shows elevated noise and DiffRecon yields ringing artifacts, AdaDiff achieves high-fidelity reconstructions with clear tissue depiction.

\begin{table}[t]
\centering
\captionsetup{justification=justified,singlelinecheck=false}
\caption{Performance of AdaDiff for varying number of diffusion steps, $T/k$. The prescribed $T/k$ for the main experiments is marked in bold font; results are listed at J=1000 iterations. FID, LPIPS, PSNR, SSIM are reported across the validation set for R=4x in IXI.}
\resizebox{0.6\columnwidth}{!}{%
\begin{tabular}{ll|c|c|c|c|c|}
\cline{2-6}
                                            & \multicolumn{1}{|c|}{}   & \multicolumn{1}{c|}{$T/k=4$} & \multicolumn{1}{c|}{$T/k=\mathbf{8}$} & \multicolumn{1}{c|}{$T/k=16$} & \multicolumn{1}{c|}{$T/k=32$} \\
 \hline
\multicolumn{1}{|c|}{\multirow{4}{*}{ T\SB{1} }} & FID  & 38.33& 31.99& 32.99& 32.78 \\ \cline{2-6}
\multicolumn{1}{|c|}{} & LPIPS  & 4.61 & 4.27 & 4.20 & 4.31  \\ \cline{2-6}
\multicolumn{1}{|c|}{} & PSNR  & 40.56 & 40.89 & 41.01 & 40.93  \\ \cline{2-6}
\multicolumn{1}{|c|}{} & SSIM  & 98.79 & 98.88 & 98.92 & 98.89  \\ \cline{2-6} \hline
\multicolumn{1}{|c|}{\multirow{4}{*}{ T\SB{2} }} & FID  & 37.75& 30.75& 30.18& 28.92 \\ \cline{2-6}
\multicolumn{1}{|c|}{} & LPIPS  & 12.80 & 12.09 & 11.81 & 11.82  \\ \cline{2-6}
\multicolumn{1}{|c|}{} & PSNR  & 40.44 & 40.93 & 40.97 & 40.97  \\ \cline{2-6}
\multicolumn{1}{|c|}{} & SSIM  & 98.20 & 98.57 & 98.70 & 98.68  \\ \cline{2-6} \hline
\multicolumn{1}{|c|}{\multirow{4}{*}{ PD }} & FID  & 38.41& 32.04& 32.92& 30.67 \\ \cline{2-6}
\multicolumn{1}{|c|}{} & LPIPS  & 12.92 & 12.63 & 12.59 & 12.72  \\ \cline{2-6}
\multicolumn{1}{|c|}{} & PSNR  & 40.79 & 41.23 & 41.29 & 41.21  \\ \cline{2-6}
\multicolumn{1}{|c|}{} & SSIM  & 98.41 & 98.75 & 98.85 & 98.82  \\ \cline{2-6} \hline
\end{tabular}
}
\label{tab:ablation1a}
\end{table}

\begin{table}[t]
\centering
\captionsetup{justification=justified,singlelinecheck=false}
\caption{Performance of AdaDiff for varying number of training epochs, $N_e$. The prescribed $N_e$ for the main experiments is marked in bold font; results are listed at J=1000 iterations.}
\resizebox{0.6\columnwidth}{!}{%
\begin{tabular}{ll|c|c|c|c|}
\cline{2-6}
                                            & \multicolumn{1}{|c|}{}  & \multicolumn{1}{c|}{$N_e$=\textbf{500}} & \multicolumn{1}{c|}{$N_e$=1000} & \multicolumn{1}{c|}{$N_e$=1500} & \multicolumn{1}{c|}{$N_e$=2000} \\
 \hline
\multicolumn{1}{|c|}{\multirow{4}{*}{ T\SB{1} }} & FID  & 31.99& 31.63& 29.88& 32.26 \\ \cline{2-6}
\multicolumn{1}{|c|}{} & LPIPS   & 4.27 & 4.20 & 4.22 & 4.34  \\ \cline{2-6}
\multicolumn{1}{|c|}{} & PSNR   & 40.89 & 40.91 & 40.88 & 40.89  \\ \cline{2-6}
\multicolumn{1}{|c|}{} & SSIM   & 98.88 & 98.88 & 98.87 & 98.88  \\ \cline{2-6} \hline
\multicolumn{1}{|c|}{\multirow{4}{*}{ T\SB{2} }} & FID  & 30.75& 29.65& 30.46& 29.96 \\ \cline{2-6}
\multicolumn{1}{|c|}{} & LPIPS   & 12.09 & 12.09 & 12.18 & 12.16  \\ \cline{2-6}
\multicolumn{1}{|c|}{} & PSNR   & 40.93 & 40.84 & 40.68 & 40.65  \\ \cline{2-6}
\multicolumn{1}{|c|}{} & SSIM   & 98.57 & 98.55 & 98.45 & 98.42  \\ \cline{2-6} \hline
\multicolumn{1}{|c|}{\multirow{4}{*}{ PD }} & FID  & 32.04& 30.41& 30.0& 29.97 \\ \cline{2-6}
\multicolumn{1}{|c|}{} & LPIPS   & 12.63 & 12.59 & 12.54 & 12.51  \\ \cline{2-6}
\multicolumn{1}{|c|}{} & PSNR  & 41.23 & 41.19 & 41.16 & 41.12  \\ \cline{2-6}
\multicolumn{1}{|c|}{} & SSIM  & 98.75 & 98.73 & 98.68 & 98.67  \\ \cline{2-6} \hline
\end{tabular}
}
\label{tab:ablation1b}
\end{table}

\subsection{Ablation Studies}
We conducted a series of ablation studies to demonstrate the main elements in AdaDiff. First, we examined the effects of number of diffusion steps, and number of training epochs on the fidelity of images synthesized by the diffusion prior during reconstruction. Table~\ref{tab:ablation1a} lists results for varying number of diffusion steps $T/k$, and Table \ref{tab:ablation1b} lists results for varying number of epochs $N_e$ across the validation set. In general, modest improvements in FID, LPIPS, PSNR and SSIM are observed with increasing $T/k$, yet the benefits in all metrics are marginal beyond $T/k=8$. The results are more intermixed for $N_e$, with FID and LPIPS showing a degree of degradation towards high $N_e$. That said, the observed differences across $N_e$ between 500 to 2000 are relatively minute for all metrics. Therefore, these results suggest that the cross-validated values of $T/k$ and $N_e$ yield near-optimal performance. 

Next, we assessed the importance of adaptive normalization of feature maps within the generator. To do this, we built a variant model that removed the latent variables $z$ to perform non-adaptive normalization (w/o $z$). Table \ref{tab:ablation1d} lists performance metrics for AdaDiff and the variant. AdaDiff outperforms the variant across reconstruction tasks, except for T\SB{1} where `w/o $z$' yields similar SSIM. This result indicates the importance of using random latents for adaptive normalization in AdaDiff.

Lastly, we examined the importance of the prior adaptation phase, the rapid diffusion phase, and the use of an adversarial mapper in reverse diffusion steps. For this purpose, we built a variant with a static diffusion prior that omitted the prior adaptation phase (w/o adapt.), a variant with an untrained diffusion prior (w/o train.), and a variant with a non-adversarial mapper by ablating the discriminator in AdaDiff and replacing the adversarial loss with a pixel-wise $\ell_1$ loss (w/o adv.). The untrained variant omitted the rapid diffusion phase to start the reconstruction with a randomly initialized generator analogously to the deep image prior method \citep{DIP}. The non-adversarial variant performed prior adaptation on a rapid diffusion prior, unlike vanilla diffusion models that interleave reverse diffusion mappings and data-consistency projections based on a slow diffusion process. Table \ref{tab:ablation2} lists performance metrics for AdaDiff and variant models at J=500 and 1000 iterations. AdaDiff outperforms all variants across reconstruction tasks (p$<$0.05 for LPIPS, PSNR, SSIM), except for T\SB{1} at J=1000 where the non-adversarial variant yields similar PSNR, SSIM. Among the examined components, the prior adaptation phase has the largest contribution to reconstruction performance as AdaDiff attains the most substantial improvement levels over the static variant. On average, AdaDiff improves FID by 363.54 ($164.97\%$ change), LPIPS by 55.43 ($146.00\%$), PSNR by 10.72dB, SSIM by 20.09\% over the static variant. In theory, the untrained and non-adversarial variants that undergo prior adaptation can converge onto an equivalent solution to AdaDiff given a very large number of iterations, as they share the same generator architecture. In practice, however, the adversarial diffusion prior in AdaDiff enables prior adaptation to start at a more favorable point and reach high performance levels in fewer iterations. AdaDiff improves FID by 18.95 ($39.42\%$ change), LPIPS by 1.59 ($14.40\%$), PSNR by 0.94dB, SSIM by 0.48\% over the untrained variant; and it improves FID by 11.02 ($24.98\%$), LPIPS by 1.74 ($15.65\%$), PSNR by 0.42dB, SSIM by 0.29\% over the non-adversarial variant. These results indicate that the rapid diffusion phase and the adversarial mapper also have important contributions to reconstruction performance, albeit at relatively modest levels. 

\begin{table}[t]
\centering
\captionsetup{justification=justified,singlelinecheck=false}
\caption{Performance of AdaDiff and a variant that omits random latent variables (w/o z). Results listed at J=1000 iterations.}
\resizebox{0.3\columnwidth}{!}{%
\begin{tabular}{ll|c|c|c|}
\cline{2-4}
                                            & \multicolumn{1}{|c|}{} & \multicolumn{1}{c|}{w/o z} & \multicolumn{1}{c|}{AdaDiff} \\
\hline
\multicolumn{1}{|c|}{\multirow{4}{*}{T\SB{1}}} & FID & 34.05 & 31.99       \\ \cline{2-4} 
\multicolumn{1}{|c|}{} & LPIPS & 4.47  & 4.27     \\ \cline{2-4} 
\multicolumn{1}{|c|}{} & PSNR & 40.84  & 40.89      \\ \cline{2-4} 
\multicolumn{1}{|c|}{} & SSIM & 98.87  & 98.88      \\ \hline
\multicolumn{1}{|c|}{\multirow{4}{*}{T\SB{2}}} & FID & 31.39  & 30.75       \\ \cline{2-4} 
\multicolumn{1}{|c|}{} & LPIPS & 12.16  & 12.09     \\ \cline{2-4} 
\multicolumn{1}{|c|}{} & PSNR & 40.83  & 40.93      \\ \cline{2-4} 
\multicolumn{1}{|c|}{} & SSIM & 98.52  & 98.57      \\ \hline
\multicolumn{1}{|c|}{\multirow{4}{*}{PD}} & FID & 32.34  & 32.04       \\ \cline{2-4} 
\multicolumn{1}{|c|}{} & LPIPS & 12.72  & 12.63     \\ \cline{2-4} 
\multicolumn{1}{|c|}{} & PSNR & 41.17  & 41.23      \\ \cline{2-4} 
\multicolumn{1}{|c|}{} & SSIM & 98.69  & 98.75      \\ \hline
\end{tabular}
}
\label{tab:ablation1d}
\end{table}

\begin{table}[t]
\centering
\captionsetup{justification=justified, singlelinecheck=false}
\caption{Performance of AdaDiff and ablated variants. A variant omitting the prior adaptation phase (w/o adapt.), a variant with an untrained prior (w/o train.), and a variant with a non-adversarial mapper (w/o adv.) were considered. Zero-filled (ZF) reconstruction results are included as reference.}
\resizebox{\columnwidth}{!}{%
\begin{tabular}{ll|c|c|c|c|c|c|c|c|c|}
\cline{3-10}
& & \multicolumn{1}{c|}{\multirow{2}{*}{ZF}} & \multicolumn{1}{c|}{\multirow{2}{*}{w/o adapt.}} & \multicolumn{3}{c|}{J=500 iterations}  & \multicolumn{3}{c|}{J=1000 iterations}  \\ 
\cline{5-10}
                                            &  &    &  & \multicolumn{1}{c|}{w/o train.} & \multicolumn{1}{c|}{w/o adv.} & \multicolumn{1}{c|}{AdaDiff} & \multicolumn{1}{c|}{w/o train.} & \multicolumn{1}{c|}{w/o adv.} & \multicolumn{1}{c|}{AdaDiff} \\ \hline
\multicolumn{1}{|c|}{\multirow{4}{*}{ T\SB{1} }} & FID  & 273.00 & 406.78 & 64.26& 64.70& 44.59& 38.88& 36.35& 31.99 \\ \cline{2-10}
\multicolumn{1}{|c|}{} & LPIPS  & 40.27 & 49.87  & 7.64& 8.74& 5.45& 5.27 & 4.89 & 4.27  \\ \cline{2-10}
\multicolumn{1}{|c|}{} & PSNR  & 32.55 & 30.85  & 39.03 & 39.65 & 40.23 & 40.33 & 40.90 & 40.89  \\ \cline{2-10}
\multicolumn{1}{|c|}{} & SSIM  & 88.80 & 84.99 & 98.06 & 98.53 & 98.71 & 98.62 & 98.90 & 98.88  \\ \cline{2-10} \hline
\multicolumn{1}{|c|}{\multirow{4}{*}{ T\SB{2} }} & FID  & 280.57 & 375.95& 65.25& 62.48& 46.64& 48.24& 34.14& 30.75 \\ \cline{2-10}
\multicolumn{1}{|c|}{} & LPIPS  & 53.64 & 76.31  & 15.07& 15.75& 13.49& 13.32 & 12.64 & 12.09  \\ \cline{2-10}
\multicolumn{1}{|c|}{} & PSNR  & 31.64 &  29.04  & 39.08 & 39.59 & 40.20 & 40.15 & 40.86 & 40.93  \\ \cline{2-10}
\multicolumn{1}{|c|}{} & SSIM  & 81.16 & 73.87  & 97.41 & 97.54 & 98.05 & 98.21 & 98.46 & 98.57  \\ \cline{2-10} \hline
\multicolumn{1}{|c|}{\multirow{4}{*}{ PD }} & FID  & 298.05 & 423.69& 73.84& 63.63& 45.60& 54.81& 36.40& 32.04 \\ \cline{2-10}
\multicolumn{1}{|c|}{} & LPIPS  & 53.64 & 70.92 & 15.59& 16.85& 13.69& 14.27 & 13.23 & 12.63  \\ \cline{2-10}
\multicolumn{1}{|c|}{} & PSNR  & 32.12 & 29.89 &  39.33 & 39.54 & 40.45 & 40.48 & 40.93 & 41.23  \\ \cline{2-10}
\multicolumn{1}{|c|}{} & SSIM  & 81.94 & 76.18 & 97.67 & 97.51 & 98.27 & 98.48 & 98.54 & 98.75  \\ \cline{2-10} \hline
\end{tabular}
}
\label{tab:ablation2}
\end{table}

Notable improvements in AdaDiff's performance are apparent in Table \ref{tab:ablation2} when J is increased from 500 to 1000. While prescribing J$>$1000 further elevates PSNR and SSIM slightly, differences in perceptual quality metrics (FID, LPIPS) and visual appearance between reconstructions at J$>$1000 and J=1000 become indiscernible (unreported). Thus, J=1000 offers a decent trade-off between reconstruction time and image quality. We also find that the static variant has suboptimal performance metrics compared to ZF and DDPM reconstructions. Both DDPM and the static variant inject data-consistency projections in between reverse diffusion steps for image reconstruction. This results in a compromise between a solution that carries realistic features of high-quality MR images based on the diffusion prior, and a solution that is anatomically consistent with the subject’s acquired k-space data based on the imaging operator. Since performance assessments involve comparisons between reconstructed and corresponding ground-truth images, they primarily reflect the anatomical consistency of reconstructions. Note that DDPM uses 1000 small steps gradually integrated with an equivalent number of data-consistency projections resulting in enhanced consistency to acquired data, and ZF natively satisfies full consistency to acquired data. In contrast, the static variant only uses 8 large steps inherently limiting consistency to acquired data and the anatomical consistency between reconstructed and ground-truth images.

\subsection{Computation Times}
A practical concern regarding MRI reconstruction methods involves training and inference times. Table \ref{tab:inferencetime} lists the computation times for competing methods, along with the static, untrained and non-adversarial variants of AdaDiff (at J=1000). In general, conditional models have shorter training times than unconditional models, and GAN models have shorter training times than diffusion models. Among diffusion-based methods, AdaDiff and the static variant have the longest training times due to the introduction of adversarial components. Note that LORAKS and the untrained variant have no training overhead. Meanwhile, LORAKS, conditional models and the non-adapted variant have notably shorter inference times than unconditional models in general. Among unconditional methods, GAN\SB{prior}, AdaDiff and its adapted variants (w/o train., w/o adv.) have comparable inference times as they all involve inference optimization procedures.

\begin{table*}[t]
\centering
\captionsetup{justification=justified,singlelinecheck=false}
\caption{Training and inference times in seconds per cross-section for reconstructions at R=4x in IXI.}
\resizebox{1.2\columnwidth}{!}{%
\begin{tabular}{l|c|c|c|c|c|c|c|c|c|c|}
\cline{2-11}
                                & LORAKS & rGAN  & MoDL & GAN\SB{prior} & DDPM & DiffRecon & w/o adapt.  & w/o train. & w/o adv. & AdaDiff \\ \hline
\multicolumn{1}{|c|}{Training} & --    & 4.0  & 26.5  & 53.8           & 15.5 & 77.3     & 131.3 & -- & 87.5 & 131.3   \\ \hline
\multicolumn{1}{|c|}{Inference} & 3.0    & 0.03 & 0.05 & 129.6          & 57.5 &  12.0    & 0.4 & 131.0  &  131.4 &  131.4   \\ \hline
\end{tabular}
}
\label{tab:inferencetime}
\end{table*}

\section{Discussion}
The proposed AdaDiff method was demonstrated against conditional and unconditional baselines for MRI reconstruction. Within-domain tasks were considered with matching acceleration rate and image distribution across the training-test sets. Several cross-domain tasks were also examined with mismatched acceleration rates, mismatched sampling trajectories, mismatched number of coils, or mismatched image distribution. We find that AdaDiff offers improved reliability against domain shifts in the imaging operator against conditional models, and against domain shifts in the MR image distribution against all competing models. Importantly, the adaptation procedure in AdaDiff notably improves reconstruction performance over competing diffusion models based on static priors for both within- and cross-domain scenarios. Yet, it remains important future work to examine reliability against broader changes in anatomy such as generalization across different body parts, and other changes in the imaging operator such as generalization across Cartesian versus non-Cartesian trajectories and across different coil arrays.   

AdaDiff learns an image prior that generates an initial reconstruction via diffusion sampling, and then adapts the prior to the test subject with an inference optimization. Despite AdaDiff's rapid diffusion process, prior adaptation naturally elevates run times over conditional models that recover images in a single forward pass. While AdaDiff is closer to regular diffusion models with iterative image sampling, it still yields relatively longer inference due to backward passes involved in prior adaptation. As expected, AdaDiff has similar inference time to the GAN-based prior adaptation method that employs a similar inference optimization. Another practical concern regarding computational complexity is memory load during inference. Among competing methods, conditional models and regular diffusion models that only leverage forward passes have relatively limited memory load. In contrast, prior adaptation methods including AdaDiff involve both forward and backward passes through the network, so they introduce additional memory load to store model gradients. To improve practicality of methods that use inference optimization, efficiency can be increased by sharing optimized model parameters across spatially proximate cross-sections within an MRI volume \citep{korkmaz2022unsupervised}, or by parallel computations on multiple GPUs. Here, we adopted the Adam algorithm observed to perform well in prior adaptation for consistency between training-inference procedures and among competing methods. Note that \cite{Knoll2019inverseGANs} originally implemented GAN\SB{prior} based on the iPALM algorithm \citep{iPALM}. It remains important future work to systematically investigate the relative benefits of different algorithms in prior adaptation, including computational efficiency and reconstruction performance.

Acceleration techniques have recently been considered to speed up the characteristically slow sampling process in regular diffusion models. An elegant approach is to initiate sampling with the image obtained from a separate reconstruction method, including zero-filled reconstructions of undersampled k-space data \citep{chung2022cvpr}. In unreported analyses, we observed that a variant model that initiated prior adaptation with zero-filled reconstructions does not offer notable benefits in performance or inference time against AdaDiff. While employing reconstructions from learning-based methods might help accelerate prior adaptation, it also necessitates independent training of secondary reconstruction models. Another powerful approach is to train diffusion models with small step size and to rescale to large step sizes during inference \citep{peng2022}. This method shortens the sampling process, but reverse diffusion steps can potentially have suboptimal accuracy. Instead, AdaDiff implements reverse diffusion over large step sizes via an adversarial mapper for improved accuracy. That said, combining the adversarial mapper in AdaDiff with the abovementioned acceleration approaches might offer further benefits.  

Several recent studies have proposed adaptation of image priors for MRI reconstruction. A group of methods reconstruct with untrained priors that map low-dimensional latent variables onto images \citep{Jin2019}. Convolutional architectures with randomly initialized weights are adopted for this purpose, wherein convolution operators serve to regularize synthesized images \citep{Arora2020ismrm,Ke2020ismrm,Zou2021,Darestani2021}. For inference, the untrained priors are combined with the imaging operator and adapted to enforce consistency between synthesized and acquired data. While performant MRI reconstruction has been reported with this approach, markedly longer inference optimization is typically required to intersect the image set reflecting the untrained prior with the image set reflecting the imaging operator \citep{Darestani2021}. An alternative group of methods instead employ priors pre-trained on MR images to provide an improved initialization point for adaptation. Previous studies in this group have predominantly proposed priors based on GAN models that implicitly characterize the MR image distribution \citep{Knoll2019inverseGANs,korkmaz2022unsupervised}, while some rely on patch-based auto-encoder models \citep{Konukoglu2019} or convolutional models \citep{aggarwal2021,Han2018}. Our work differs from recent methods based on untrained priors in that AdaDiff leverages a prior pre-trained on high-quality coil-combined MR images to improve efficiency during inference optimization. It also differs from methods with pre-trained priors since AdaDiff leverages a novel diffusion prior to improve fidelity of image samples. 

In theory, prior adaptation can be performed based on regular diffusion models instead of the adversarial diffusion model considered here. However, regular diffusion models might elicit several limitations in the context of prior adaptation. Note that generation of the initial reconstruction with regular diffusion models involves image sampling across hundreds of reverse diffusion steps interleaved with data-consistency projections \citep{chung2022cvpr}. Thus, computing the initial reconstruction with regular diffusion models requires comparable inference time to the prior adaptation stage (e.g., see DDPM and AdaDiff in Table \ref{tab:inferencetime}), significantly elevating the overall computational burden. Furthermore, the initial reconstruction based on hundreds of data-consistency projections naturally yields an enhanced match to acquired k-space data. In turn, regular diffusion models already yield low data-consistency loss, significantly limiting the added benefit that can be achieved via prior adaptation. To improve prior adaptation with regular diffusion models, data-consistency projections might be partly omitted during the initial reconstruction stage to ensure a reasonably high level of data-consistency loss. It remains future work to examine the optimal training and inference procedures for prior adaptation with regular diffusion models.

Reconstruction performance for AdaDiff might be improved through several lines of technical development. First, all reported models were trained by pooling acquisitions across multiple distinct contrasts in each dataset. The trained models were then used for independently reconstructing individual MRI contrasts. Performance might be improved by training separate models on each contrast, at the expense of computational burden \citep{Dar2017}. When a multi-contrast accelerated MRI protocol is available in each subject, joint reconstruction models can also be used to exploit structural correlations among contrasts to improve performance \citep{rgan,Polakjointvvn2020,gaillochet2020joint,xuan2022multi}. For AdaDiff, this would involve training of a multi-contrast diffusion prior. Alternatively, contrast type can be provided as side information to maintain specificity to individual contrasts in a unified model \citep{liu2022undersampled,dalmaz2022one}. Second, cycle-consistent learning strategies can be adopted to alleviate the dependence of AdaDiff on datasets comprising fully-sampled acquisitions \citep{Quan2018c,oh2020,ozbey2022unsupervised}. Here, AdaDiff was implemented to perform prior adaptation by optimizing the generator parameters at the final time step for computational efficiency. In theory, performing prior adaptation on the entire set of diffusion steps could improve reconstruction performance. In practice, however, inference optimization over multiple diffusion steps requires computation and storage of gradients across all steps. In turn, this would substantially elevate the memory load and inference time for AdaDiff, limiting practical utility.

The primary focus of the current study was on a learning strategy to improve generalization in diffusion-based MRI reconstruction. Thus, we adopted convolutional generator and discriminator architectures reported to offer high performance in previous studies on generative modeling \citep{song2020score,DiffNvidia,StyleGAN2}. Further work is warranted to assess the contributions of various design elements in the employed architectures to AdaDiff's performance. Future studies should also be conducted to explore the utility of alternative architectures such as transformer backbones \citep{dalmaz2021resvit,transms,guo2022reconformer}, and the influence of different normalization layers on model performance. To represent complex MRI data, we used separate network channels for the real and imaginary components following common practice in learning-based MRI reconstruction \citep{Schlemper2017,KikiNet,Variatonal_end2end,MoDl,ADMM-CSNET}. Recent studies suggest that complex-valued network operations might offer benefits particularly in phase-oriented reconstruction tasks \citep{complexnet1,kustner2020cinenet,complexnet3,complexnet4}. It remains important future work to explore the potential benefits of adopting complex-valued operations in AdaDiff.

\section{Conclusion}
In this study, we introduced the first prior adaptation method based on diffusion modeling for MRI reconstruction. AdaDiff leverages an adversarial mapper for reverse diffusion that enables efficient image generation in few steps. During inference, an initial reconstruction is obtained via rapid projection through the trained diffusion prior. The final reconstruction is then computed by further adapting the prior to the test subject. Compared against state-of-the-art baselines, AdaDiff performs competitively in within-domain tasks, and achieves superior reconstructions in cross-domain tasks. Therefore, AdaDiff holds great promise for high-performance MRI reconstruction. 

\section*{Acknowledgments}
This study was supported in part by a TUBITAK BIDEB scholarship awarded to A. Gungor, by a TUBITAK BIDEB scholarship awarded to S. Ozturk, and by a TUBITAK 1001 Research Grant (121E488), a TUBA GEBIP 2015 fellowship, and a BAGEP 2017 fellowship awarded to T. \c{C}ukur.

\bibliographystyle{model2-names}\biboptions{authoryear}
\bibliography{Papers}

\end{document}